\documentclass[aps,prd,preprint,groupedaddress,longbibliography,nofootinbib]{revtex4-1}

\usepackage{amsmath}
\usepackage{graphicx}
\newcommand{\vev}[1]{\left\langle #1 \right\rangle}
\newcommand{\vvev}[1]{\vev{\kern-0.3em\left\langle #1
    \right\rangle\kern-0.3em}}

\newcommand{\lb}{\left\lbrace}
  \newcommand{\rb}{\right\rbrace}

\newcommand{\fs}[1]{\hbox{$#1$\kern-0.5em\raise0.3ex\hbox{/}}}

\newcommand{\nt}{\notag}

\begin{document}


\title{Background dependent cutoff for Wilson actions}


\author{Carlo Pagani} \email{cpagani@uni-mainz.de}
\affiliation{Institute f\"{u}r Physik (WA THEP)
  Johannes-Gutenberg-Universit\"{a}t\\ Staudingerweg 7, 55099 Mainz,
  Germany}

\author{Hidenori Sonoda} \email[Visiting Research Associate
till October 2024,~]{h-sonoda@pobox.com} \affiliation{Department of
  Physics and Astronomy, The University of Iowa, Iowa City, Iowa
  52242, USA}


\date{\today}

\begin{abstract}
We study the application of the background field method in the framework of the exact renormalization group (ERG).
By considering the case of a scalar field theory,
we provide a detailed discussion of the properties satisfied by a background dependent Wilson action.
We show how the Ward identity associated with background field shifts for the Wilson action is
related to that for the 1PI Wilson action, or effective average action.
Moreover, we discuss the ERG equations in the dimensionless framework and emphasize the role of the Ward identity in this setting.
We give two examples: the Gaussian fixed point for the real scalar
theory and the large $N$ limit of the linear sigma model.
\end{abstract}


\maketitle


\section{Introduction}

The use of background fields in quantum field theory (QFT) is essential
for several applications, such as QFT in curved space or in a strong
magnetic field. Moreover, the background field method \cite{DeWitt:1964mxt,Abbott:1980hw}
offers a way to quantize gauge theory that preserves a form of gauge
invariance, i.e., the so-called background gauge symmetry. 

Within the framework of Wilsonian renormalization group, the use of
background fields has been widely exploited within the effective
average action formalism
\cite{Reuter:1993kw,Reuter:1996cp,Freire:2000bq}, the main
applications being gauge theories and quantum gravity, see
e.g.~\cite{Gies:2006wv,Reuter_Saueressig_2019,Dupuis:2020fhh}.
However, within the Wilson action formalism, the background field
method has been little explored. In this paper we study in details the
introduction of a background field into Wilson actions.  The key point
is to carefully keep track of field independent terms in the exact
renormalization group (ERG) equation for the Wilson action.
Furthermore, in order to keep the physics intact and independent of
the background used, certain Ward identities must be satisfied. We
construct such Ward identities for the Wilson action and discuss in
detail how they are related to the analogous identities in the
effective average action formalism
\cite{Wetterich:1992yh,Reuter:1993kw,Morris:1993qb,Ellwanger:1993mw}.
In so doing, we also highlight certain aspects of these identities
that have remained uncovered so far.

In this work we focus on the case of a scalar field theory. Within the
effective average action formalism, the use of the background field
method for a scalar field theory has been studied previously
\cite{Litim:2002hj,Manrique:2009uh,Manrique:2010mq,Manrique:2010am,Bridle:2013sra,Dietz:2015owa,Labus:2016lkh}.
The case of a background dependent cutoff function was studied in
refs.~\cite{Manrique:2009uh,Manrique:2010mq,Manrique:2010am,Bridle:2013sra,Dietz:2015owa,Labus:2016lkh}
as a benchmark to discuss split symmetry in the context of the
asymptotic safety scenario for quantum gravity
\cite{Reuter_Saueressig_2019,Percacci_book_2017}.  In this paper, we
deal with both background independent and background dependent cutoff
functions. On top of constructing the Ward identities for Wilson
actions, we show how these are related to those associated with the
generating functionals of connected Green functions and of
one-particle-irreducible (1PI) functions. In this setting we discuss
the role of the Ward identities in detail.

The paper is organized as follows.  In section
\ref{sec:Wilson-action-and-background-fields} we introduce a
background field in the framework of Wilson actions and discuss the
Ward identity associated with background shifts (also known as split
symmetry Ward identity).  In section
\ref{sec:generating-functionals-and-split-symm} we extend our
discussion to the generating functionals of connected Green functions
and of 1PI functions. We show how the identities obtained for Wilson
actions are related to those obtained for these other functionals.  
In section \ref{sec:anom_dim} and \ref{sec:dimless_ERG} we introduce the anomalous dimension and the dimensionless framework, respectively. Furthermore, we discuss how the Ward identity associated with background shifts allows one to simplify the background dependent ERG.
In
section \ref{sec:checks-of-split-WI} we discuss two examples: the
Gaussian fixed point for a scalar theory and the large $N$ limit of
the linear sigma model.  We summarize our findings in section
\ref{sec:conclusions}.

Throughout the paper we work in the $d$-dimensional Euclidean space.
For integrals in momentum space and in coordinate space we use the
following short hand notation
\begin{equation}
  \int_p = \int \frac{d^d p}{(2 \pi)^d},\quad
  \int_x = \int d^d x\,,
\end{equation}
specifying just the integrated variables.

\section{Wilson action and background fields} \label{sec:Wilson-action-and-background-fields}

\subsection{ERG and background fields \label{subsec:ERG-and-background-fields}}

The Wilson action $S_{\Lambda}$ is defined as a particular solution to
the associated ERG differential equation. Given such a solution, one
may calculate the correlation functions of the theory by integrating
out the degrees of freedom below the cutoff $\Lambda$. For the sake of
simplicity, we shall work with a real scalar field and indicate the
background field by $\bar{\phi}$ and the fluctuating field by $\phi$.
The Wilson action is denoted by
$S_{\Lambda}\left[\phi,\bar{\phi}\right]$.

As a warm-up, we shall consider background independent cutoff
functions.  Let us start by considering the following ERG equation:
\begin{align}
&-\Lambda\frac{\partial}{\partial\Lambda}e^{S_{\Lambda}\left[\phi,\bar{\phi}\right]}
                \nt\\
  & =  \int_{p}\frac{\delta}{\delta\phi\left(p\right)}\lb \Biggr[\Lambda\frac{\partial\log K_{\Lambda}\left(p\right)}{\partial\Lambda}\phi\left(p\right)+\frac{1}{2}\Lambda\frac{\partial\log R_{\Lambda}\left(p\right)}{\partial\Lambda}\frac{R_{\Lambda}\left(p\right)}{K_{\Lambda}\left(p\right)^{2}}\frac{\delta}{\delta\phi\left(p\right)}\Biggr]e^{S_{\Lambda}\left[\phi,\bar{\phi}\right]}\rb\,,\label{eq:ERG-cutoff-no-background-1}
\end{align}
where $K_{\Lambda}\left(p\right)$ and $R_{\Lambda}\left(p\right)$ are
cutoff functions. The RHS of Eq.~(\ref{eq:ERG-cutoff-no-background-1})
is a total differential of the fluctuating field. This guarantees that
the partition function is independent of the cutoff $\Lambda$, as has
been well known since the pioneering works of Wilson and Wegner
\cite{Wilson:1973jj,Wegner:1972ih} and as has been recently analyzed
in detail in the Wilsonian framework in \cite{Pagani:2024gdt}.

Let us start by considering the background field method in standard
quantum field theory. The correlation functions associated with a bare
action $S_{\rm{bare}}\left[\phi\right]$ satisfy
\begin{eqnarray}
\langle\phi\left(p_{1}\right)\cdots\phi\left(p_{n}\right)\rangle_{S_{\rm{bare}}\left[\bar{\phi}+\Delta+\phi\right]} & = & \langle\left(\phi\left(p_{1}\right)-\Delta\left(p_{1}\right)\right)\cdots\left(\phi\left(p_{n}\right)-\Delta\left(p_{n}\right)\right)\rangle_{S_{\rm{bare}}\left[\bar{\phi}+\phi\right]}\,.\label{eq:standard_QFT_background_field_correlations}
\end{eqnarray}
We wish to construct a Wilson action such that the correlation functions
satisfy a relation equivalent to (\ref{eq:standard_QFT_background_field_correlations}).

To make contact with the Wilson action we consider the so-called modified
correlation function \cite{Sonoda:2015bla} and impose the relation
analogous to (\ref{eq:standard_QFT_background_field_correlations}):
\begin{equation}
  \vvev{\phi\left(p_{1}\right)\cdots\phi\left(p_{n}\right)}_{\bar{\phi}+\Delta}
  =
  \vvev{\left(\phi\left(p_{1}\right)-\Delta\left(p_{1}\right)\right)\cdots\left(\phi\left(p_{n}\right)
      -\Delta\left(p_{n}\right)\right)}_{\bar{\phi}}\,,
  \label{eq:imposing-split-no-background-in-cutoffs}  
\end{equation}
where we define
\begin{align}
&\vvev{\phi\left(p_{1}\right)\cdots\phi\left(p_{n}\right)}_{\bar{\phi}}
                   \nt\\
  & \equiv
    \prod_{i=1}^{n}\frac{1}{K_{\Lambda}\left(p_{i}\right)}\Bigr\langle\exp\left(-\frac{1}{2}\int_{p}\frac{K_{\Lambda}\left(p\right)^{2}}{R_{\Lambda}\left(p\right)}\frac{\delta^{2}}{\delta\phi\left(p\right)\delta\phi\left(-p\right)}\right)\phi\left(p_{1}\right)\cdots\phi\left(p_{n}\right)\Bigr\rangle_{S_\Lambda
    [\phi, \bar{\phi}]}\,.\label{eq:def-mod-corr-funct-1}
\end{align}

By considering the infinitesimal version of
Eq.~(\ref{eq:imposing-split-no-background-in-cutoffs}) one finds
\begin{eqnarray}
\int_{p}\Delta\left(p\right)\frac{\delta S_{\Lambda}\left[\phi,\bar{\phi}\right]}{\delta\bar{\phi}\left(p\right)} & = & \int_{p}\Delta\left(p\right)K_{\Lambda}\left(p\right)\frac{\delta S_{\Lambda}\left[\phi,\bar{\phi}\right]}{\delta\phi\left(p\right)}\,,\label{eq:WI_Wilson_action_cutoff_no_backgr}
\end{eqnarray}
which implies
\begin{eqnarray}
S_{\Lambda}\left[\phi,\bar{\phi}\right] & = & S_{\Lambda}\left[\phi+K_{\Lambda}\bar{\phi}\right]\,.\label{eq:S_Lambda-cutoff-with-no-background-1}
\end{eqnarray}
Equation (\ref{eq:S_Lambda-cutoff-with-no-background-1}) tells us that
the Wilson action keeps a simple dependence on the background:
$S_{\Lambda}\left[\phi,\bar{\phi}\right]$ is just a functional of the
combination $\phi+K_{\Lambda}\bar{\phi}$, which is reminiscent of the
split-symmetric combination $\phi+\bar{\phi}$ in the bare action
$S_0 [\phi + \bar{\phi}]$.  However, as we shall see, such a simple
relation applies only to the case where the cutoff function has no
dependence on $\bar{\phi}$.

\subsection{Introducing a background dependent cutoff function \label{subsec:Intr-background-dep-cutoff-function}}

In this section we generalize the reasoning of section
\ref{subsec:ERG-and-background-fields} by considering background
dependent cutoff functions, which we denote by
$K_{\Lambda}\left[\bar{\phi}\right]$ and
$R_{\Lambda}\left[\bar{\phi}\right]$.  This extension is needed, for
instance, if one wanted to apply the background field method to gauge
theories for which a Wilson action with manifest background gauge
invariance is desired. In order to disentangle different sources of
background dependence, we shall not consider gauge theories but work
on a real scalar field theory; in gauge theories there is an extra
complication of gauge fixing that depends on the background.

For a generic space dependent background field $\bar{\phi} (x)$ we
introduce cutoff functions
$K_{\Lambda}\left[\bar{\phi}\right]\left(x,y\right)$ and
$R_{\Lambda}\left[\bar{\phi}\right]\left(x,y\right)$ as integration
kernels.  They are real, positive, and symmetric:
\begin{equation}
  K_\Lambda [\bar{\phi}] (x,y) = K_\Lambda [\bar{\phi}] (y,x),\quad
  R_\Lambda [\bar{\phi}] (x,y) = R_\Lambda [\bar{\phi}] (y,x)
\end{equation}
For instance, we may introduce $R_\Lambda [\bar{\phi}] (x,y)$ by the
diffusion equation
\begin{subequations}
\begin{align}
  - \Lambda \partial_\Lambda R_\Lambda [\bar{\phi}] (x,y)
  &= 2 \left( - 1 + \frac{2}{\Lambda^2} \partial_x^2 + (d-2) \alpha
    \frac{1}{\Lambda^{d-2}} \bar{\phi} (x)^2 \right) R_\Lambda
    [\bar{\phi}] (x,y)\,,\\
  &=  2 \left( - 1 + \frac{2}{\Lambda^2} \partial_y^2 + (d-2) \alpha
    \frac{1}{\Lambda^{d-2}} \bar{\phi} (y)^2 \right) R_\Lambda
    [\bar{\phi}] (x,y)\,,
\end{align}
\label{RL-diffusion}
\end{subequations}
and the asymptotic condition
\begin{equation}
  \lim_{\Lambda \to +\infty} \frac{1}{\Lambda^2} R_\Lambda
  [\bar{\phi}] (x,y) = \delta (x-y)\,.
\end{equation}
The inverse $R_\Lambda^{-1} [\bar{\phi}] (x,y)$ is defined by
\begin{equation}
  \int_y R_\Lambda^{-1} [\bar{\phi}] (x,y) R_\Lambda [\bar{\phi}]
  (y,z)
  = \int_y R_\Lambda [\bar{\phi}] (x,y) R^{-1}_\Lambda [\bar{\phi}]
  (y,z) = \delta (x-z)\,.
\end{equation}
Similarly we define $K_\Lambda [\bar{\phi}] (x,y)$ and its inverse.\footnote{
Alternatively, we
may consider a differential operator defined by
\begin{eqnarray}
  \hat{R}_{\Lambda} & \equiv & \Lambda^{2}\Biggr\{\sum_{i=0}^\infty r_{i}\left(\frac{-\partial_{x}^{2}}{\Lambda^{2}}+\alpha\frac{\bar{\phi}\left(x\right)^2}{\Lambda^{d-2}}\right)^{i}\Biggr\} 
\end{eqnarray}
and the associated cutoff function by
$R_{\Lambda}\left[\bar{\phi}\right]\left(x,y\right) 
\equiv 
\hat{R}_\Lambda \delta (x-y)$,
which is symmetric by construction.
The cutoff function $K\left[\bar{\phi}\right]\left(x,y\right)$ can be defined similarly.}

Now the modified correlation functions are defined by
\begin{align}
  &\vvev{\phi (x_1) \cdots \phi (x_n)}
  \equiv \prod_{i=1}^n \int_{x'_i} K^{-1}_\Lambda [\bar{\phi}] (x_i, x'_i) \nt\\
&\quad\times  \vev{\exp \left( - \frac{1}{2} \int_{x,y}
      \left(K_\Lambda R^{-1}_\Lambda K_\Lambda\right) (x,y)
      \frac{\delta^2}{\delta \phi (x) \delta \phi (y)}\right)\, \phi
    (x'_1) \cdots \phi (x'_n)}_{S_\Lambda [\phi, \bar{\phi}]}\,,
\end{align}
where we use a condensed notation for
\begin{equation}
  \left(K_\Lambda R^{-1}_\Lambda K_\Lambda \right) (x,y)
  \equiv \int_{u,v} K_\Lambda [\bar{\phi}] (x,u) R^{-1}_\Lambda
  [\bar{\phi}] (u,v) K_\Lambda [\bar{\phi}] (v, y)\,.
  \end{equation}
  The $\Lambda$-independence of the modified correlation functions
  implies the ERG equation:
\begin{align}
-\Lambda\frac{\partial}{\partial\Lambda}e^{S_{\Lambda}\left[\phi,\bar{\phi}\right]}
  & =  \int_{x,y}\frac{\delta}{\delta\phi\left(x\right)}\Bigg\lbrace
    \Biggr[\left(\Lambda \partial_\Lambda K_{\Lambda} \cdot
    K_\Lambda^{-1}\right)\left(x,y\right)\phi\left(y\right)\nt\\ 
  &\quad + \frac{1}{2}\left(K_\Lambda R_\Lambda^{-1} \Lambda
    \partial_\Lambda R_\Lambda R_\Lambda^{-1} K_\Lambda\right)
    \left(x,y\right)\frac{\delta}{\delta\phi\left(y\right)}\Biggr]
    e^{S_{\Lambda}\left[\phi,\bar{\phi}\right]}\Bigg\rbrace\,, \label{eq:ERG-cutoff-background-1}
\end{align}
where condensed notations are used.
Let us emphasize that the terms
independent of the fluctuating field $\phi$ must be kept as they
contribute crucially to background dependent terms.

By applying the same logic as in the previous section let us impose
\begin{eqnarray}
\vvev{\phi\left(x_{1}\right)\cdots\phi\left(x_{n}\right)}_{\bar{\phi}+\Delta} & = & \vvev{\left(\phi\left(x_{1}\right)-\Delta\left(x_{1}\right)\right)\cdots\left(\phi\left(x_{n}\right)-\Delta\left(x_{n}\right)\right)}_{\bar{\phi}}\,.\label{eq:imposing-split-with-background-in-cutoffs-1}
\end{eqnarray}
in coordinate space first.
We then consider an infinitesimal shift of $\bar{\phi}$.  Denoting
\begin{align}
&  \vvev{\mathcal{O} [\phi] \phi (x_1) \cdots \phi (x_n)}_{\bar{\phi}}
  \equiv \prod_{i=1}^n \int_{x'_i} K_\Lambda [\bar{\phi}] (x_i, x'_i)\nt\\
&\qquad  \times\vev{ \mathcal{O} [\phi] \exp \left( - \frac{1}{2}
     \int_{x,y} \left(K_\Lambda R_\Lambda^{-1} K_\Lambda\right) (x,y) 
      \frac{\delta^2}{\delta \phi (x) \delta \phi (y)}\right)\,\phi
    (x'_1) \cdots \phi (x'_n)}_{S_\Lambda [\phi, \bar{\phi}]}\,,
\end{align}
where $\mathcal{O}[\phi]$ is any functional of $\phi$, and expanding
Eq.~(\ref{eq:imposing-split-with-background-in-cutoffs-1}) to first
order in $\Delta$, we obtain
\begin{align}
  &
    \int_{x,y}\Delta\left(x\right)\,K_\Lambda [\bar{\phi}] (x,y)\vev{\kern-0.5em\vev{
    \frac{\delta S_\Lambda [\phi,\bar{\phi}]}{\delta \phi (y)}\,
    \phi\left(x_{1}\right)\cdots
    \phi \left(x_{n}\right)}\kern-0.5em}_{\bar{\phi}}  \nonumber \\  
  &=  \int_x \Delta\left(x\right)\vev{\kern-0.5em\vev{\frac{\delta
    S_{\Lambda} [\phi,\bar{\phi}]}{\delta\bar{\phi}\left(x\right)}\,\phi\left(x_{1}\right)\cdots
    \phi\left(x_{n}\right)}\kern-0.5em}_{\bar{\phi}}\nonumber  \\ 
  &     \quad -\int_{x,y,z} \Delta\left(x\right)
    \lb \frac{\delta}{\delta\bar{\phi}\left(x\right)}\left(
    K_\Lambda R^{-1}_\Lambda K_\Lambda \right) (y,z)\rb\nonumber\\
  &   \qquad\times   \vev{\kern-0.5em\vev{
    e^{-S_{\Lambda}[\phi,\bar{\phi}]}\left(\frac{1}{2}\frac{\delta^{2}
    e^{S_{\Lambda}[\phi,\bar{\phi}]}}{\delta\phi\left(y\right) 
    \delta\phi\left(z\right)}
    \right)\phi\left(x_{1}\right)\cdots
    \phi\left(x_{n}\right)}\kern-0.5em}_{\bar{\phi}}\nonumber   \\ 
  & \quad  + \int_{x,y,z} \Delta\left(x\right) 
    \left(\frac{\delta K_\Lambda}{\delta \bar{\phi}
    (x)} K_\Lambda^{-1}\right) (y,z) \nt\\
  &\qquad\times \Bigg\langle\kern-0.5em\Bigg\langle e^{-S_\Lambda
    [\phi,\bar{\phi}]}
    \frac{\delta}{\delta \phi
    (y)} 
    \lb \left( \phi (z) + \int dw\, \left(K_\Lambda R_\Lambda^{-1}
    K_\Lambda\right) (y,w) \frac{\delta}{\delta \phi (w)} \right)
    e^{S_\Lambda [\phi, \bar{\phi}]} \rb \nt\\
  &\qquad\quad \times \phi (x_1) \cdots \phi
    (x_n)\Bigg\rangle\kern-0.5em\Bigg\rangle_{\bar{\phi}}\,,
\end{align}
We thus obtain
\begin{align}
  &\frac{\delta S_\Lambda [\phi, \bar{\phi}]}{\delta
  \bar{\phi} (x)}
  = \int_y K_\Lambda [\bar{\phi}] (x,y) \frac{\delta
    S_\Lambda [\phi, \bar{\phi}]}{\delta \phi (y)} \nt\\
  &\quad + \int_{y,z}
    \left(\frac{\delta K_\Lambda}{\delta \bar{\phi} (x)}
    K^{-1}_\Lambda \right) (y,z) e^{-S_\Lambda [\phi,\bar{\phi}]}
    \frac{\delta}{\delta \phi (y)}\nt\\
  &\qquad \times
    \lb \left( \phi (z) + \int_w \left(K_\Lambda R^{-1}_\Lambda
    K_\Lambda\right) (z,w) \frac{\delta}{\delta \phi (w)} \right)
    e^{S_\Lambda [\phi, \bar{\phi}]} \rb\nt\\
   &\quad + \frac{1}{2} \int_{y,z} \frac{\delta}{\delta \bar{\phi}
     (x)}
     \left(K_\Lambda R^{-1}_\Lambda K_\Lambda\right) (y,z)\,
     e^{-S_\Lambda [\phi,\bar{\phi}]} \frac{\delta^2}{\delta \phi (y)
     \delta \phi (z)} e^{S_\Lambda [\phi, \bar{\phi}]}\,.
  \label{eq:WI_Wilson_action_cutoff_with_backgr2}
\end{align}
Equation (\ref{eq:WI_Wilson_action_cutoff_with_backgr2}) has the same
logical status as Eq.~(\ref{eq:WI_Wilson_action_cutoff_no_backgr}).
The two additional terms in
(\ref{eq:WI_Wilson_action_cutoff_with_backgr2}) are entirely due to the
background dependence of the cutoff functions.

\subsection{Spin variables}

In ref.~\cite{Pagani:2024gdt} so called ``spin variables'' were
introduced. The main property of these variables is to make the Wilson
action very simple in the limit $\Lambda\rightarrow0$; we refer to
\cite{Pagani:2024gdt} for a detailed discussion on this point. It is
straightforward to introduce such variables in the presence of a
background. We define spin variables by
\begin{equation}
  \sigma\left(x\right)  \equiv \int_y \left(R_\Lambda K^{-1}_\Lambda\right)
                                (x,y) \phi (y)
                                = \int_{y,z} R_\Lambda [\bar{\phi}]
                                (x,z) K^{-1}_\Lambda [\bar{\phi}]
                                (z,y) \phi (y)\,.
\end{equation}
Keeping track of the (background dependent) Jacobian, we obtain the
Wilson action in terms of the spin variables as
\cite{Pagani:2024gdt}
\begin{equation}
e^{S_{\Lambda} \left[\sigma,\bar{\phi}\right]} 
                                                    =\int[d\phi']
\exp \left( - \frac{1}{2} \int_{x,y} \phi' (x)
                                                    \left(K_\Lambda^{-1}
                                                      R_\Lambda K_\Lambda^{-1}\right)
                                                    (x,y) \phi'
                                                    (y)\right)
                                                  \times e^{S_\Lambda
                                                    [\phi, \bar{\phi}]}\,,
\end{equation}
where we apply a particular convention for the integration measure
such that
\begin{eqnarray}
\int[d\sigma]\exp\left(-\frac{1}{2}\int_x \sigma(x)^2\right) & = & 1\,.
\end{eqnarray}
Note that we denote the Wilson action in terms of the spin
variables as $S_{\Lambda}\left[\sigma,\bar{\phi}\right]$ rather than
something like $\tilde{S}_\Lambda$.  Its ERG differential equation
reads
\begin{eqnarray}
  -\Lambda\frac{\partial}{\partial\Lambda}e^{S_{\Lambda}\left[\sigma,\bar{\phi}\right]}
  & = & \int_{x,y} \left(\Lambda
        \frac{\partial \sqrt{R}_\Lambda}{\partial \Lambda} \,R_\Lambda^{-\frac{1}{2}}\right) (x,y)
        \frac{\delta}{\delta\sigma(x)}
        \left[\left(\sigma(y)+\frac{\delta}{\delta\sigma(y)}\right)
        e^{S_{\Lambda}\left[\sigma,\bar{\phi}\right]}\right]\,.
\end{eqnarray}
In the limit $\Lambda\rightarrow0$ one finds
$S_{\Lambda}\left[\sigma,\bar{\phi}\right]=-\frac{1}{2}\int_{p}\sigma\left(p\right)\sigma\left(-p\right)+F
[\bar{\phi}]$, where $F$ is independent of $\sigma$.  See
\cite{Pagani:2024gdt}.

With spin variables, the modified correlation functions are given by
\begin{align}
  \vvev{\phi(x_{1})\cdots\phi(x_{n})}_{\bar{\phi}}
  & \equiv \prod_{i=1}^n \int_{x'_i} R_\Lambda^{-\frac{1}{2}} [\bar{\phi}] (x_i, x'_i)\nt\\
      &\quad \times \vev{\exp \left( - \frac{1}{2} \int_x
        \frac{\delta^2}{\delta \sigma (x) \delta \sigma (x)}\right)\,
        \sigma (x'_1) \cdots \sigma (x'_n)}_{S_\Lambda [\sigma, \bar{\phi}]}\,,
\end{align}
where the symmetric kernel $\sqrt{R}_\Lambda [\bar{\phi}]$ satisfies
\begin{equation}
  \int_y \sqrt{R}_\Lambda [\bar{\phi}] (x,y) \sqrt{R}_\Lambda
  [\bar{\phi}] (y,z) = R_\Lambda [\bar{\phi}] (x-z)\,,
\end{equation}
and $R^{-\frac{1}{2}}_\Lambda [\bar{\phi}]$ is its inverse.  If
$R_\Lambda [\bar{\phi}]$ satisfies (\ref{RL-diffusion}),
$\sqrt{R}_\Lambda [\bar{\phi}]$ satisfies
\begin{equation}
  - \Lambda \partial_\Lambda \sqrt{R}_\Lambda (x,y) = \left( - 1 +
    \frac{2}{\Lambda^2} \partial_x^2 + (d-2) \alpha 
    \frac{1}{\Lambda^{d-2}} \bar{\phi} (x)^2 \right) \sqrt{R}_\Lambda (x,y)\,.
\end{equation}
Imposing Eq.~(\ref{eq:imposing-split-with-background-in-cutoffs-1}),
we obtain a somewhat simplified version of
(\ref{eq:WI_Wilson_action_cutoff_with_backgr2}) as
\begin{align}
  \frac{\delta}{\delta\bar{\phi}\left(x\right)} e^{S_\Lambda [\sigma,\bar{\phi}]}
  &= \int_y \sqrt{R}_\Lambda [\bar{\phi}] (x,y) \frac{\delta
    }{\delta \sigma (y)} e^{S_\Lambda [\sigma,\bar{\phi}]}  \nt\\
&\quad - \int_{y,z} \left(\frac{\delta \sqrt{R}_\Lambda}{\delta
            \bar{\phi} (x)} R^{-\frac{1}{2}}_\Lambda \right)(y,z)
\frac{\delta}{\delta \sigma (y)} \lb \left(\sigma (z) +
    \frac{\delta}{\delta \sigma (z)}\right) e^{S_\Lambda [\sigma, \bar{\phi}]}\rb\,.
\label{eq:split-WI-for-sigma-1}
\end{align}
(This is obtained from (\ref{eq:WI_Wilson_action_cutoff_with_backgr2})
by setting $K_\Lambda = \sqrt{R}_\Lambda$.)

\subsection{Representation in terms of a bare action \label{subsec:Representation-in-terms-S0}}

So far we have avoided referring to a bare action as much as possible.
However, it is sometimes handy to consider a bare action even if it
may not exist. In particular, with the help of a bare action, it is
possible to give functional integral representations of the various
Wilson actions introduced such as $S_\Lambda [\phi, \bar{\phi}]$ and
$S_\Lambda [\sigma, \bar{\phi}]$. Let
\begin{align}
&  e^{S_{\Lambda}\left[\phi, \bar{\phi}\right]}  =
                                      \frac{1}{\int [d
                                      \phi^{\prime\prime}]\,\exp\left[-\frac{1}{2}
                \int_{x,y} \phi^{\prime\prime}\left(x\right)
                                      \left(K_\Lambda^{-1} R_\Lambda
                                      K_\Lambda^{-1}\right)
                                      (x,y) 
                                      \phi^{\prime\prime} (y)\right]}
\times \int [d \phi^{\prime}]\,\exp\Bigg[S_{\rm{bare}}\left[\phi^{\prime} + \bar{\phi}\right]\nt\\
&\,  -\frac{1}{2} \int_{x,y} R_\Lambda [\bar{\phi}] (x,y) \left(
  \phi' (x) - \int_u K_\Lambda^{-1}(x,u) \phi (u)\right)
  \left( \phi' (y) - \int_v K_\Lambda^{-1} (y,v) \phi (v)\right)
  \Bigg]\,,
  \label{eq:Wilson_action_representation_via_Sbare}
\end{align}
where $S_{\rm{bare}}$ is a bare action.  The Wilson action defined
formally by (\ref{eq:Wilson_action_representation_via_Sbare}) has the
correct $\Lambda$-dependence. The representation
(\ref{eq:Wilson_action_representation_via_Sbare}) is sometimes easier
to use, for example in the derivation of split Ward identity
(\ref{eq:WI_Wilson_action_cutoff_with_backgr2}).

In full similarity, one can introduce the following representation
for $S_{\Lambda}[\sigma, \bar{\phi}]$ via a bare action:
\begin{align}
&e^{S_{\Lambda}\left[\sigma, \bar{\phi}\right]}  =  \int
                [d\phi^{\prime}]\,\exp\Bigg[S_{\rm{bare}}\left[\phi^{\prime}
                +\bar{\phi}\right]\nt\\
  &\quad -\frac{1}{2}\int_x \left(\int_y
    \sqrt{R}_{\Lambda}\left[\bar{\phi}\right]\left(x,
    y\right) \phi^{\prime}\left(y\right)-\sigma\left(x\right)\right)\left(
    \int_z \sqrt{R}_{\Lambda}\left[\bar{\phi}\right]\left(x, z
    \right) \phi^{\prime}\left(z\right)-\sigma\left(x\right)\right)\Bigg]\,.\label{eq:S-Sigma_action_representation_via_Sbare}
\end{align}
We can easily derive the split Ward identity
(\ref{eq:split-WI-for-sigma-1}) from this.  Therefore, we have two
possible ways of deriving the split Ward identity for background
shifts. On the one hand, we may proceed to derive the identity
starting from the representation
(\ref{eq:S-Sigma_action_representation_via_Sbare}).  On the other
hand, we may impose the identity
(\ref{eq:imposing-split-with-background-in-cutoffs-1}) on the
correlation functions of the model and observe that this is realized
only if the same Ward identity is satisfied. The latter viewpoint
offers the advantage of working with a finite cutoff throughout, not
relying on the existence of a bare action.

\section{Generating functionals and split symmetry}
\label{sec:generating-functionals-and-split-symm}

\subsection{Generating functional for connected correlation functions}

We define the generating functional for connected correlation functions
by
\begin{eqnarray}
W_{\Lambda}\left[J,\bar{\phi}\right] & \equiv &
                                                S_{\Lambda}\left[\sigma,\bar{\phi}\right]
                                                +\frac{1}{2}\int_x
                                                \lb\sigma\left(x\right)\rb^2\label{eq:def_WL_via_Ssigma} 
\end{eqnarray}
with
\begin{eqnarray}
J\left(x\right) & \equiv &
\int_y \sqrt{R}_{\Lambda}\left[\bar{\phi}\right]\left(x,y\right)\,
                           \sigma\left(y\right)\,.
\end{eqnarray}
In analogy with section \ref{subsec:Representation-in-terms-S0}, we
can give a simple representation of $W_{\Lambda}$ in terms of a bare
action:
\begin{align}
  &  e^{W_{\Lambda}\left[J,\bar{\phi}\right]} \nt\\
  & = \int    [d\phi]\,\exp\left[S_{\rm{bare}}\left[\bar{\phi}+\phi\right]-\frac{1}{2}
    \int_{x,y} \phi\left(x\right) R_{\Lambda}\left[\bar{\phi}\right]\left(x,y\right)
    \phi\left(y\right)  +\int_x  J\left(x\right)\phi\left(x\right)\right]\,.
\label{eq:functional_int_representation_WL} 
\end{align}
Equation (\ref{eq:functional_int_representation_WL}) reproduces the
correct $\Lambda$-dependence
\begin{eqnarray}
-\Lambda\frac{\partial}{\partial\Lambda}e^{W_{\Lambda}\left[J,\bar{\phi}\right]}
  & = & \frac{1}{2}\int_{x,y}
        \Lambda\frac{\partial R_{\Lambda}\left[\bar{\phi}\right]\left(x,y\right)}{\partial\Lambda}\frac{\delta^{2}}{\delta J\left(x\right)\delta J\left(y\right)}e^{W_{\Lambda}\left[J,\bar{\phi}\right]}\,,
\end{eqnarray}
which can be derived directly from the definition
(\ref{eq:def_WL_via_Ssigma}).

At this point, it is natural to introduce also the functional $W_{\Lambda}^{\prime}\left[J,\bar{\phi}\right]$:
\begin{equation}
e^{W_{\Lambda}^{\prime}\left[J,\bar{\phi}\right]} = \int
 [d\phi]\,\exp\left[S_{\rm{bare}}\left[\phi\right]-\frac{1}{2}\int_{x,y}
                                                        \phi\left(x\right)
                                                        R_{\Lambda}[\bar{\phi}]\left(x,y\right)
                                                        \phi\left(y\right)
                                                        +\int_x
                                                        J\left(x\right)\phi\left(x\right)\right]\,,
                                                      \label{eq:functional_int_representation_WprimeL}
\end{equation}
where the field $\phi$ in the bare action is not shifted.  By
definition, the background dependence in
$W_{\Lambda}^{\prime}\left[J,\bar{\phi}\right]$ comes only from the
background dependence of the cutoff function.  Though we have
introduced $W_{\Lambda}^{\prime}\left[J,\bar{\phi}\right]$ by
Eq.~(\ref{eq:functional_int_representation_WprimeL}), we could have
equally well used the definition (\ref{eq:def_WL_via_Ssigma}) using
$S_{\Lambda}^{\prime} [\sigma, \bar{\phi}]$ with no shift in the field
variables.  However, we find it technically easier to work with the
integral representations (\ref{eq:functional_int_representation_WL})
and (\ref{eq:functional_int_representation_WprimeL}).  It is
straightforward to derive the relation between the two functionals:
\begin{equation}
  W_\Lambda [J, \bar{\phi}]
  = W'_\Lambda \left[ J + R_\Lambda \bar{\phi}, \bar{\phi}\right] -
  \int_x J(x) \bar{\phi} (x) - \frac{1}{2} \int_{x,y} \bar{\phi}
  (x) R_\Lambda [\bar{\phi}] (x,y) \bar{\phi} (y)\,,\label{eq:WL-vs-WLprime}
\end{equation}
where
\begin{equation}
\left(  R_\Lambda \bar{\phi}\right) (x) \equiv \int_y R_\Lambda
[\bar{\phi}] (x,y) \bar{\phi} (y)\,.
\end{equation}

\subsection{Split Ward identities for $W_{\Lambda}$ and $W_{\Lambda}^{\prime}$}

It is straightforward to derive the Ward identities associated with
background shifts. One can proceed in two ways. On the one hand, one
may consider a background variation at fixed source $J$ of the definition
(\ref{eq:def_WL_via_Ssigma}). By using equation (\ref{eq:split-WI-for-sigma-1}),
one obtains
\begin{align}
\frac{\delta
  W_{\Lambda}\left[J,\bar{\phi}\right]}{\delta\bar{\phi}\left(x\right)}
  &=  -\frac{1}{2}\int_{y,z} \frac{\delta
    R_{\Lambda}\left[\bar{\phi}\right]\left(y,z\right)}{\delta\bar{\phi}\left(x\right)}\Biggr[\frac{\delta 
    W_{\Lambda}\left[J,\bar{\phi}\right]}{\delta
    J\left(y\right)}\frac{\delta
    W_{\Lambda}\left[J,\bar{\phi}\right]}{\delta
    J\left(z\right)}+\frac{\delta^{2}W_{\Lambda}\left[J,\bar{\phi}\right]}{\delta
    J\left(y\right)\delta J\left(z\right)}\Biggr]\nonumber \\ 
 & \quad + \int_y R_{\Lambda}\left[\bar{\phi}\right]\left(x,
   y\right)\frac{\delta W_{\Lambda}\left[J,\bar{\phi}\right]}{\delta
   J\left(y\right)}-J\left(x\right)\,.\label{eq:split_sym_WI_WL} 
\end{align}
On the other hand, Eq.~(\ref{eq:split_sym_WI_WL}) may be also readily
derived from the functional integral representation in terms of a bare
action, i.e., Eq.~(\ref{eq:functional_int_representation_WL}).  Once
again, despite obtaining the same result, let us emphasize that in the
first approach one works by keeping at each step an explicit cutoff
and imposes an appropriate property on the correlation functions.  In
the second approach, one assumes that a bare action with a certain
property exists and derives the Ward identity as a
consequence. Therefore, the logic behind the two approaches differs
somewhat.

The split symmetry Ward identity for $W_{\Lambda}^{\prime}$ takes
the following form:
\begin{equation}
\frac{\delta
  W_{\Lambda}^{\prime}\left[J,\bar{\phi}\right]}{\delta\bar{\phi}\left(x\right)}
   =  -\frac{1}{2}\int_{y,z} \frac{\delta
    R_{\Lambda}\left[\bar{\phi}\right]\left(y,z\right)}{\delta\bar{\phi}\left(x\right)}\Biggr[\frac{\delta
    W'_{\Lambda}\left[J,\bar{\phi}\right]}{\delta
    J\left(y\right)}\frac{\delta
    W'_{\Lambda}\left[J,\bar{\phi}\right]}{\delta
    J\left(z\right)}+\frac{\delta^{2}W'_{\Lambda}\left[J,\bar{\phi}\right]}{\delta
    J\left(y\right)\delta
    J\left(z\right)}\Biggr]\,.\label{eq:split_sym_WI_WprimeL} 
\end{equation}
The terms on the RHS of (\ref{eq:split_sym_WI_WprimeL}) are due only
to the background dependence of the cutoff. 

\subsection{Generating functional of 1PI vertices and split symmetry}

We define the 1PI Wilson action $\Gamma_{\Lambda}$, or the effective
average action, by the Legendre transformation
\begin{eqnarray}
\Gamma_{\Lambda}\left[\Phi,\bar{\phi}\right]-\frac{1}{2}\int_{x,y}
\phi\left(x\right)R_{\Lambda}\left[\bar{\phi}\right]\left(x,y\right)\phi\left(y\right)
  & = & W_{\Lambda}\left[\Phi,\bar{\phi}\right]-\int_x  J\left(x\right)\Phi\left(x\right)\,,
\end{eqnarray}
where
\begin{eqnarray}
\Phi\left(x\right) & \equiv & \frac{\delta W_{\Lambda}\left[\Phi,\bar{\phi}\right]}{\delta J\left(x\right)}\,.
\end{eqnarray}
It follows that
\begin{eqnarray}
J\left(x\right) & = & \int_y
                      R_{\Lambda}\left[\bar{\phi}\right]\left(x,y\right)\Phi\left(y\right)-\frac{\delta\Gamma_{\Lambda}\left[\Phi,\bar{\phi}\right]}{\delta\Phi\left(x\right)}\,.
                      \label{J-for-GammaL}
\end{eqnarray}

Similarly, we define $\Gamma_{\Lambda}^{\prime}$ by
\begin{eqnarray}
\Gamma_{\Lambda}^{\prime}\left[\Phi,\bar{\phi}\right]-\frac{1}{2}\int_{x,y}
  \Phi\left(x\right)R_{\Lambda}\left[\bar{\phi}\right]\left(x,y\right)\Phi\left(y\right)
  & = & W_{\Lambda}^{\prime}\left[\Phi,\bar{\phi}\right]
        - \int_x J\left(x\right)\Phi\left(y\right)\,,
\end{eqnarray}
where
\begin{eqnarray}
\Phi\left(x\right) & \equiv & \frac{\delta W_{\Lambda}^{\prime}\left[\Phi,\bar{\phi}\right]}{\delta J\left(x\right)}\,.
\end{eqnarray}
It follows that
\begin{eqnarray}
J\left(x\right) & = & \int_y
                      R_{\Lambda}\left[\bar{\phi}\right]\left(x,y\right)\Phi\left(y\right)-\frac{\delta\Gamma_{\Lambda}^{\prime}\left[\Phi,\bar{\phi}\right]}{\delta\Phi\left(x\right)}\,.
                      \label{J-for-GammaLprime}
\end{eqnarray}

The 1PI actions $\Gamma_{\Lambda}$ and $\Gamma_{\Lambda}^{\prime}$ are
related: using (\ref{eq:WL-vs-WLprime}), it is straightforward to
obtain
\begin{eqnarray}
\Gamma_{\Lambda}\left[\Phi,\bar{\phi}\right] & = & \Gamma_{\Lambda}^{\prime}\left[\Phi+\bar{\phi},\bar{\phi}\right]\,,\label{eq:relation-GammaL-GammaPrimeL}
\end{eqnarray}
or equivalently
\begin{equation}
  \Gamma_\Lambda [\Phi - \bar{\phi}, \bar{\phi}] = \Gamma'_\Lambda
  [\Phi, \bar{\phi}]\,.
\end{equation}
In the limit $\Lambda \to 0+$, $\Gamma'_\Lambda$ reduces to the
effective action
\begin{equation}
  \lim_{\Lambda \to 0+} \Gamma'_\Lambda [\Phi, \bar{\phi}] =
  \Gamma_{\mathrm{eff}} [\Phi]
\end{equation}
which is independent of the background $\bar{\phi}$.  
Choosing $\Phi = \bar{\phi}$, we obtain
the well known result
\begin{equation}
  \lim_{\Lambda\rightarrow 0+} \Gamma_{\Lambda}\left[0,\bar{\phi}\right]
  = \lim_{\Lambda \to 0+} \Gamma_{\Lambda}^{\prime}\left[\bar{\phi},\bar{\phi}\right]  =  \Gamma_{\mathrm{eff}}\left[\bar{\phi}\right]\,.
\end{equation}

It is straightforward to derive the ERG equations and split Ward
identities for the 1PI Wilson actions.  We only state the results.
The ERG equations are given by
\begin{subequations}
  \begin{align}
    - \Lambda \partial_\Lambda \Gamma_\Lambda [\Phi, \bar{\phi}]
    &= \frac{1}{2} \int_{x,y} \Lambda \partial_\Lambda R_\Lambda
      [\bar{\phi}] (x,y)\, \frac{\delta^2 W_\Lambda [J,
      \bar{\phi}]}{\delta J(x) \delta J(y)}\,,\label{ERG-for-GammaL}\\
     - \Lambda \partial_\Lambda \Gamma'_\Lambda [\Phi, \bar{\phi}]
    &= \frac{1}{2} \int_{x,y} \Lambda \partial_\Lambda R_\Lambda
      [\bar{\phi}] (x,y)\, \frac{\delta^2 W'_\Lambda [J,
      \bar{\phi}]}{\delta J(x) \delta J(y)}\,,\label{ERG-for-GammaLprime}
  \end{align}
\end{subequations}
and the split Ward identities by
\begin{subequations}
\begin{align}
  \frac{\delta\Gamma_{\Lambda}\left[\Phi,\bar{\phi}\right]}{\delta\bar{\phi}\left(z\right)}
  & =  \frac{\delta\Gamma_{\Lambda}\left[\Phi,\bar{\phi}\right]}{\delta\Phi\left(z\right)}-\frac{1}{2}\int_{x,y}\frac{\delta R_{\Lambda}\left[\bar{\phi}\right]\left(x,y\right)}{\delta\bar{\phi}\left(z\right)}\frac{\delta^{2}W_{\Lambda}\left[J,\bar{\phi}\right]}{\delta J\left(x\right)\delta J\left(y\right)}\,,\label{splitWI-for-GammaL}\\
  \frac{\delta\Gamma_{\Lambda}^{\prime}\left[\Phi,\bar{\phi}\right]}{\delta\bar{\phi}\left(z\right)}
  & =  -\frac{1}{2}\int_{x,y}\frac{\delta R_{\Lambda}\left[\bar{\phi}\right]\left(x,y\right)}{\delta\bar{\phi}\left(z\right)}\frac{\delta^{2}W_{\Lambda}^{\prime}\left[J,\bar{\phi}\right]}{\delta J\left(x\right)\delta J\left(y\right)}\,,\label{splitWI-for-GammaLprime}
\end{align}
\end{subequations}
where the quadratic differentials of $W_\Lambda, W'_\Lambda$ are
obtained from
\begin{subequations}
\begin{align}
\int_{z}\left(R_{\Lambda}\left[\bar{\phi}\right]\left(x,z\right)-\frac{\delta^{2}\Gamma_{\Lambda}\left[\Phi,\bar{\phi}\right]}{\delta\Phi\left(x\right)\delta\Phi\left(z\right)}\right)\frac{\delta^{2}W_{\Lambda}\left[J,\bar{\phi}\right]}{\delta
  J\left(z\right)\delta J\left(y\right)}
  & = \delta\left(x-y\right)\,,\\
\int_{z}\left(R_{\Lambda}\left[\bar{\phi}\right]\left(x,z\right)-\frac{\delta^{2}\Gamma_{\Lambda}^{\prime}\left[\Phi,\bar{\phi}\right]}{\delta\Phi\left(x\right)\delta\Phi\left(z\right)}\right)\frac{\delta^{2}W_{\Lambda}^{\prime}\left[J,\bar{\phi}\right]}{\delta
  J\left(z\right)\delta J\left(y\right)}
  &= \delta\left(x-y\right)\,.
\end{align}
\end{subequations}
Note that for the same $\Phi (x)$, we have two different variables
$J (x)$ depending on whether we consider
$\Gamma_\Lambda [\Phi, \bar{\phi}]$ or
$\Gamma'_\Lambda [\Phi, \bar{\phi}]$.  For
$\Gamma_\Lambda [\Phi, \bar{\phi}]$, we have Eq.~(\ref{J-for-GammaL})
giving $J (x)$.  For $\Gamma'_\Lambda [\Phi, \bar{\phi}]$, we have
Eq.~(\ref{J-for-GammaLprime}) giving $J (x)$.  Those two $J(x)$'s are
different.  $J(x)$ given by Eq.~(\ref{J-for-GammaL}) for
$\Gamma_\Lambda [\Phi, \bar{\phi}]$ is related to $J' (x)$ given by
Eq.~(\ref{J-for-GammaLprime}) for
$\Gamma'_\Lambda [\Phi + \bar{\phi}, \bar{\phi}]$ as follows:
\begin{align}
  J' (x)
  &\equiv \int_y R_\Lambda [\bar{\phi}] (x,y) \left(\Phi (y) +
    \bar{\phi} (y)\right) - \frac{\delta \Gamma'_\Lambda [\Phi +
    \bar{\phi}, \bar{\phi}]}{\delta \Phi (x)}\nt\\
  &= \int_y R_\Lambda [\bar{\phi}] (x,y) \Phi (y) - \frac{\delta
    \Gamma_\Lambda [\Phi, \bar{\phi}]}{\delta \Phi (x)} + \int_y
    R_\Lambda [\bar{\phi}] (x,y) \bar{\phi} (y)\nt\\
  &= J(x) + \int_y R_\Lambda [\bar{\phi}] (x, y) \bar{\phi} (y)\,.
\end{align}

\section{Introducing the anomamlous dimension \label{sec:anom_dim}}

In this section we introduce an anomalous dimension of the real scalar
field that affects all the functionals considered in this paper.
Except for the first subsection, we will not elaborate much about the
necessary changes due to the anomalous dimension; we will mostly
enumerate the results.

\subsection{Anomalous dimension for the Wilson action}

To introduce an anomalous dimension for the scalar field $\phi$, let
us define
\begin{align}
  \vvev{\phi\left(x_{1}\right)\cdots\phi\left(x_{n}\right)}_{\Lambda}^{\prime}
  &\equiv  
                \prod_{i=1}^{n}\int_{x_{i}^{\prime}}R_{\Lambda}^{-1/2}\left(x_{i},x_{i}^{\prime}\right)\nt\\
  &\quad \times \left\langle\exp\left(-\frac{1}{2}\int_{x}\frac{\delta^{2}}{\delta\sigma\left(x\right)\delta\sigma\left(x\right)}\right)\sigma\left(x_{1}^{\prime}\right)\cdots\sigma\left(x_{n}^{\prime}\right)\right\rangle_{S_{\Lambda}^{\prime}\left[\sigma,\bar{\phi}\right]} \,,
\end{align}
and require the $\Lambda$-dependence
\begin{eqnarray}
\vvev{\phi\left(x_{1}\right)\cdots\phi\left(x_{n}\right)}_{\Lambda}^{\prime}
  & \propto & z_{\Lambda}^{n}\,. 
\end{eqnarray}
The anomalous dimension is defined by
\begin{eqnarray}
\gamma_{\Lambda} & \equiv & -z_{\Lambda}^{-1}\Lambda\partial_{\Lambda}z_{\Lambda}\,.
\end{eqnarray}
This implies that the Wilson action $S_{\Lambda}^{\prime}\left[\sigma,\bar{\phi}\right]$
satisfies the following ERG equation:
\begin{eqnarray}
-\Lambda\partial_{\Lambda}e^{S_{\Lambda}^{\prime}\left[\sigma,\bar{\phi}\right]}
  & = & \int_{x,y}\left\{
        -\gamma_{\Lambda}\delta\left(x-y\right)+\left(\Lambda
        \frac{\partial \sqrt{R_{\Lambda}}}{\partial \Lambda} R_\Lambda^{-1/2}\right)\left(x,y\right)\right\} 
\nonumber \\
 &  & \times\frac{\delta}{\delta\sigma\left(x\right)}\left\{ \left(\sigma\left(y\right)+\frac{\delta}{\delta\sigma\left(y\right)}\right)\right\} e^{S_{\Lambda}^{\prime}\left[\sigma,\bar{\phi}\right]}\,.
\end{eqnarray}
The same ERG equation can be obtained from the following integral
representation:
\begin{eqnarray}
e^{S_{\Lambda}^{\prime}\left[\sigma,\bar{\phi}\right]} & = & \int\left[d\phi\right]\,\exp\left[S_{{\rm bare}}\left[\phi\right]-\frac{1}{2}\int_{x}\left(\sigma\left(x\right)-z_{\Lambda}\int_{y}\sqrt{R_{\Lambda}}\left[\bar{\phi}\right]\left(x,y\right)\phi\left(y\right)\right)^{2}\right]\,.
\end{eqnarray}

Next, let us define
\begin{eqnarray}
S_{\Lambda}\left[\sigma,\bar{\phi}\right] & \equiv & S_{\Lambda}^{\prime}\left[\sigma+z_{\Lambda}\sqrt{R_{\Lambda}}\bar{\phi},\bar{\phi}\right]\,,
\end{eqnarray}
where
$\left(\sqrt{R_{\Lambda}}\bar{\phi}\right)\left(x\right)\equiv\int_{y}\sqrt{R_{\Lambda}}
[\bar{\phi}] \left(x,y\right)\bar{\phi}\left(y\right)$.
The modified correlation functions given by
\begin{align}
  \vvev{\phi\left(x_{1}\right)\cdots\phi\left(x_{n}\right)}_{\Lambda}
  &\equiv
    \prod_{i=1}^{n}\int_{x_{i}^{\prime}}R_{\Lambda}^{-1/2}\left(x_{i},x_{i}^{\prime}\right)\nt\\
  &\quad \times \Bigr\langle\exp\left(-\frac{1}{2}\int_{x}\frac{\delta^{2}}{\delta\sigma\left(x\right)\delta\sigma\left(x\right)}\right)\sigma\left(x_{1}^{\prime}\right)\cdots\sigma\left(x_{n}^{\prime}\right)\Bigr\rangle_{S_{\Lambda}\left[\sigma,\bar{\phi}\right]}
\end{align}
satisfy the following relation:
\begin{eqnarray}
  \vvev{\left(\phi\left(x_{1}\right)+z_{\Lambda}\bar{\phi}\left(x_{1}\right)\right)\cdots\left(\phi\left(x_{n}\right)+z_{\Lambda}\bar{\phi}\left(x_{n}\right)\right)}_{\Lambda}
  & \equiv &
             \vvev{\phi\left(x_{1}\right)\cdots\phi\left(x_{n}\right)}_{\Lambda}^{\prime}\,.
\end{eqnarray}
Finally, we can give the following integral representation to
$S_{\Lambda}\left[\sigma,\bar{\phi}\right]$:
\begin{eqnarray}
e^{S_{\Lambda}\left[\sigma,\bar{\phi}\right]} & = & \int\left[d\phi\right]\,\exp\left[S_{{\rm bare}}\left[\bar{\phi}+\phi\right]-\frac{1}{2}\int_{x}\left(\sigma\left(x\right)-z_{\Lambda}\int_{y}\sqrt{R_{\Lambda}}\left[\bar{\phi}\right]\left(x,y\right)\phi\left(y\right)\right)^{2}\right]
.
\end{eqnarray}

The Ward identities associated with background shifts for $S_{\Lambda}^{\prime}\left[\sigma,\bar{\phi}\right]$
and $S_{\Lambda}\left[\sigma,\bar{\phi}\right]$ take now the following
form:
\begin{subequations}
\begin{equation}
\frac{\delta}{\delta\bar{\phi}\left(x\right)}e^{S_{\Lambda}^{\prime}\left[\sigma,\bar{\phi}\right]}
=  -\int_{y, z}\left(\frac{\delta\sqrt{R_{\Lambda}}}{\delta\bar{\phi}\left(x\right)}\sqrt{R_{\Lambda}}^{-1}\right)\left(y,z\right)\frac{\delta}{\delta\sigma\left(y\right)}\left\{ \left(\sigma\left(z\right)+\frac{\delta}{\delta\sigma\left(z\right)}\right)e^{S_{\Lambda}^{\prime}\left[\sigma,\bar{\phi}\right]}\right\} 
\end{equation}
and
\begin{align}
\frac{\delta}{\delta\bar{\phi}\left(x\right)}e^{S_{\Lambda}\left[\sigma,\bar{\phi}\right]}
  &= z_{\Lambda}\int_{y}\sqrt{R_{\Lambda}}\left(x,y\right)\frac{\delta}{\delta\sigma\left(y\right)}e^{S_{\Lambda}\left[\sigma,\bar{\phi}\right]}\nonumber\\
 & \quad  -\int_{y,z}\left(\frac{\delta\sqrt{R_{\Lambda}}}{\delta\bar{\phi}\left(x\right)}\sqrt{R_{\Lambda}}^{-1}\right)\left(y,z\right)\frac{\delta}{\delta\sigma\left(y\right)}\left\{ \left(\sigma\left(z\right)+\frac{\delta}{\delta\sigma\left(z\right)}\right)e^{S_{\Lambda}\left[\sigma,\bar{\phi}\right]}\right\} \,,
\end{align}
\end{subequations}
respectively.

\subsection{Anomalous dimension for $W_{\Lambda}^{\prime}$ and $W_{\Lambda}$}

We define
\begin{eqnarray}
J\left(x\right) & \equiv & \int_{y}\sqrt{R_{\Lambda}}\left[\bar{\phi}\right]\left(x,y\right)\sigma\left(y\right)
\end{eqnarray}
and
\begin{subequations}
\begin{eqnarray}
W_{\Lambda}^{\prime}\left[J,\bar{\phi}\right] & \equiv & S_{\Lambda}^{\prime}\left[\sigma,\bar{\phi}\right]+\frac{1}{2}\int_{x}\sigma\left(x\right)^{2}\,,\\
W_{\Lambda}\left[J,\bar{\phi}\right] & \equiv & S_{\Lambda}\left[\sigma,\bar{\phi}\right]+\frac{1}{2}\int_{x}\sigma\left(x\right)^{2}\,.
\end{eqnarray}
\end{subequations}
Alternatively, we can give the following integral representations to
$W_{\Lambda}^{\prime}\left[J,\bar{\phi}\right]$ and
$W_{\Lambda}\left[J,\bar{\phi}\right]$:
\begin{subequations}
\begin{align}
e^{W_{\Lambda}^{\prime}\left[J,\bar{\phi}\right]} &\equiv \int\left[d\phi\right]\,\exp\left[S_{{\rm bare}}\left[\phi\right]+\int_{x}J\left(x\right)z_{\Lambda}\phi\left(x\right)-\frac{z_{\Lambda}^{2}}{2}\int_{x,y}\phi\left(x\right)R_{\Lambda}\left[\bar{\phi}\right]\left(x,y\right)\phi\left(y\right)\right]\label{eq:functional-repr-Wpr}\\
e^{W_{\Lambda}\left[J,\bar{\phi}\right]} &\equiv
                                                    \int\left[d\phi\right]\,\exp\left[S_{{\rm bare}}\left[\bar{\phi}+\phi\right]+\int_{x}J\left(x\right)z_{\Lambda}\phi\left(x\right)
 -\frac{z_{\Lambda}^{2}}{2}\int_{x,y}\phi\left(x\right)R_{\Lambda}\left[\bar{\phi}\right]\left(x,y\right)\phi\left(y\right)\right]\,.\label{eq:functional-repr-W}
\end{align}
\end{subequations}

By using either the relation
$S_{\Lambda}\left[\sigma,\bar{\phi}\right]=S_{\Lambda}^{\prime}\left[\sigma+z_{\Lambda}R_{\Lambda}\bar{\phi},\bar{\phi}\right]$
or the above integral representations of
$W_{\Lambda}^{\prime}\left[J,\bar{\phi}\right]$ and
$W_{\Lambda}\left[J,\bar{\phi}\right]$ one obtains
\begin{equation}
W'_{\Lambda}\left[J,\bar{\phi}\right]  =
W_{\Lambda}\left[J+z_{\Lambda}R_{\Lambda}\bar{\phi},\bar{\phi}\right]
- z_\Lambda \int_{x}J\left(x\right) \bar{\phi}\left(x\right)-\frac{1}{2}z_\Lambda^2
\int_{x,y}\bar{\phi}\left(x\right)R_{\Lambda}\left[\bar{\phi}\right]\left(x,y\right)\bar{\phi}\left(y\right)\,.\label{eq:relation-WprimeLambda-WLambda}
\end{equation}
The Ward identities are obtained as
\begin{subequations}
  \label{eq:WLprime-WL-WI}
\begin{eqnarray}
\frac{\delta}{\delta\bar{\phi}\left(x\right)}e^{W_{\Lambda}^{\prime}\left[J,\bar{\phi}\right]} & = & -\frac{1}{2}\int_{y,z}\frac{\delta R_{\Lambda}\left[\bar{\phi}\right]\left(y,z\right)}{\delta\bar{\phi}\left(x\right)}\frac{\delta^{2}}{\delta J\left(y\right)\delta J\left(z\right)}e^{W_{\Lambda}^{\prime}\left[J,\bar{\phi}\right]}\,,\\
\frac{\delta}{\delta\bar{\phi}\left(x\right)}e^{W_{\Lambda}\left[J,\bar{\phi}\right]} & = & \Biggr(-z_{\Lambda}J\left(x\right)+z_{\Lambda}\int_{y}R_{\Lambda}\left[\bar{\phi}\right]\left(x,y\right)\frac{\delta}{\delta J\left(y\right)}
\nonumber \\
 &  & \quad -\frac{1}{2}\int_{y,z}\frac{\delta R_{\Lambda}\left[\bar{\phi}\right]\left(y,z\right)}{\delta\bar{\phi}\left(x\right)}\frac{\delta^{2}}{\delta J\left(y\right)\delta J\left(z\right)}\Biggr)e^{W_{\Lambda}^{\prime}\left[J,\bar{\phi}\right]}\,.
\end{eqnarray}
\end{subequations}

\subsection{Anomalous dimension for $\Gamma_{\Lambda}^{\prime}$ and $\Gamma_{\Lambda}$}

We consider the following two modified Legendre transforms
\begin{subequations}
\begin{eqnarray}
\Gamma_{\Lambda}^{\prime}\left[\Phi,\bar{\phi}\right]-\frac{1}{2}\int_{x,y}\Phi\left(x\right)R_{\Lambda}\left[\bar{\phi}\right]\left(x,y\right)\Phi\left(y\right) & = & W_{\Lambda}^{\prime}\left[J,\bar{\phi}\right]-\int_{x}J\left(x\right)\Phi\left(x\right)\,,
\end{eqnarray}
where
\begin{eqnarray}
\Phi\left(x\right) & \equiv & \frac{\delta W_{\Lambda}^{\prime}\left[J,\bar{\phi}\right]}{\delta J\left(x\right)}
\end{eqnarray}
\end{subequations}
and
\begin{subequations}
\begin{eqnarray}
\Gamma_{\Lambda}\left[\Phi,\bar{\phi}\right]-\frac{1}{2}\int_{x,y}\Phi\left(x\right)R_{\Lambda}\left[\bar{\phi}\right]\left(x,y\right)\Phi\left(y\right) & = & W_{\Lambda}\left[J,\bar{\phi}\right]-\int_{x}J\left(x\right)\Phi\left(x\right)\,,
\end{eqnarray}
where
\begin{eqnarray}
\Phi\left(x\right) & \equiv & \frac{\delta W_{\Lambda}\left[J,\bar{\phi}\right]}{\delta J\left(x\right)}\,.
\end{eqnarray}
\end{subequations}
By using the relation (\ref{eq:relation-WprimeLambda-WLambda}) between
$W_{\Lambda}^{\prime}$ and $W_{\Lambda}$ one finds
\begin{eqnarray}
\Gamma_{\Lambda}\left[\Phi,\bar{\phi}\right] & = &
                                                   \Gamma_{\Lambda}^{\prime}\left[\Phi+z_{\Lambda}\bar{\phi},\bar{\phi}\right]\,.
                                                   \label{eq:relation-Gammaprime-Gamma}
\end{eqnarray}

The $\Lambda$-dependence of $\Gamma_{\Lambda}^{\prime}$ and $\Gamma_{\Lambda}$
is governed by the following ERG equations:
\begin{subequations}
\begin{eqnarray}
-\Lambda\partial_{\Lambda}\Gamma_{\Lambda}^{\prime}\left[\Phi,\bar{\phi}\right] & = & -\gamma_{\Lambda}\int_{x}\frac{\delta\Gamma_{\Lambda}^{\prime}\left[\Phi,\bar{\phi}\right]}{\delta\Phi\left(x\right)}\Phi\left(x\right)
\nonumber \\
 &  & +\frac{1}{2}\int_{x,y}\left(\Lambda\partial_{\Lambda}R_{\Lambda}\left[\bar{\phi}\right]\left(x,y\right)-2\gamma_{\Lambda}R_{\Lambda}\left[\bar{\phi}\right]\left(x,y\right)\right)\frac{\delta^{2}W_{\Lambda}^{\prime}\left[J,\bar{\phi}\right]}{\delta J\left(x\right)\delta J\left(y\right)}\,,
\end{eqnarray}
where
\begin{eqnarray}
\int_{y}\frac{\delta^{2}W_{\Lambda}^{\prime}\left[J,\bar{\phi}\right]}{\delta J\left(x\right)\delta J\left(y\right)}\left(R_{\Lambda}\left[\bar{\phi}\right]\left(y,z\right)-\frac{\delta^{2}\Gamma_{\Lambda}^{\prime}\left[\Phi,\bar{\phi}\right]}{\delta\Phi\left(y\right)\delta\Phi\left(z\right)}\right) & = & \delta\left(x-z\right)\,,
\end{eqnarray}
\end{subequations}
and
\begin{subequations}
\begin{eqnarray}
-\Lambda\partial_{\Lambda}\Gamma_{\Lambda}\left[\Phi,\bar{\phi}\right] & = & -\gamma_{\Lambda}\int_{x}\frac{\delta\Gamma_{\Lambda}\left[\Phi,\bar{\phi}\right]}{\delta\Phi\left(x\right)}\Phi\left(x\right)
\nonumber \\
 &  & +\frac{1}{2}\int_{x,y}\left(\Lambda\partial_{\Lambda}R_{\Lambda}\left[\bar{\phi}\right]\left(x,y\right)-2\gamma_{\Lambda}R_{\Lambda}\left[\bar{\phi}\right]\left(x,y\right)\right)\frac{\delta^{2}W_{\Lambda}\left[J,\bar{\phi}\right]}{\delta J\left(x\right)\delta J\left(y\right)}\,,
\end{eqnarray}
where
\begin{eqnarray}
\int_{y}\frac{\delta^{2}W_{\Lambda}\left[J,\bar{\phi}\right]}{\delta J\left(x\right)\delta J\left(y\right)}\left(R_{\Lambda}\left[\bar{\phi}\right]\left(y,z\right)-\frac{\delta^{2}\Gamma_{\Lambda}\left[\Phi,\bar{\phi}\right]}{\delta\Phi\left(y\right)\delta\Phi\left(z\right)}\right) & = & \delta\left(x-z\right)\,.
\end{eqnarray}
\end{subequations}

By using the Ward identities (\ref{eq:WLprime-WL-WI}) associated with background shifts
one obtains
\begin{subequations}
  \label{eq:GammaLprime-GammaL-WI}
\begin{eqnarray}
\frac{\delta}{\delta\bar{\phi}\left(x\right)}\Gamma_{\Lambda}^{\prime}\left[\Phi,\bar{\phi}\right] & = & -\frac{1}{2}\int_{yz}\frac{\delta R_{\Lambda}\left[\bar{\phi}\right]\left(y,z\right)}{\delta\bar{\phi}\left(x\right)}\frac{\delta^{2}W_{\Lambda}^{\prime}\left[J,\bar{\phi}\right]}{\delta J\left(y\right)\delta J\left(z\right)}\\
\frac{\delta}{\delta\bar{\phi}\left(x\right)}\Gamma_{\Lambda}\left[\Phi,\bar{\phi}\right] & = & z_{\Lambda}\frac{\delta\Gamma_{\Lambda}\left[\Phi,\bar{\phi}\right]}{\delta\Phi\left(x\right)}-\frac{1}{2}\int_{yz}\frac{\delta R_{\Lambda}\left[\bar{\phi}\right]\left(y,z\right)}{\delta\bar{\phi}\left(x\right)}\frac{\delta^{2}W_{\Lambda}^{\prime}\left[J,\bar{\phi}\right]}{\delta J\left(y\right)\delta J\left(z\right)}\,,
\end{eqnarray}
\end{subequations}
which are consistent with the relation
(\ref{eq:relation-Gammaprime-Gamma}).

\section{Dimensionless ERG equation \label{sec:dimless_ERG}}

\subsection{Dimensionless framework}

We define the dimensionless fields by
\begin{eqnarray}
 &
   \tilde{\sigma}\left(x\right)\equiv\Lambda^{-d/2}\sigma\left(x/\Lambda\right)\,,\,\tilde{\Phi}\left(x\right)\equiv\Lambda^{-\frac{d-2}{2}}\Phi\left(x/\Lambda\right)\,,\,\tilde{\bar{\phi}}\left(x\right)\equiv\Lambda^{-\frac{d-2}{2}}\bar{\phi}\left(x/\Lambda\right)\,,
   \label{eq:dimless-fields}
\end{eqnarray}
where the tildes are used to distinguish the dimensionless fields.
The action and the 1PI action in the dimensionless framework are then
defined by
\begin{align}
\tilde{S}_{t}^{\prime}\left[\tilde{\sigma},\tilde{\bar{\phi}}\right]=S_{\Lambda}^{\prime}\left[\sigma,\bar{\phi}\right]\,, &  & \tilde{S}_{t}\left[\tilde{\sigma},\tilde{\bar{\phi}}\right]=S_{\Lambda}\left[\sigma,\bar{\phi}\right]\,,
\end{align}
and
\begin{align}
\tilde{\Gamma}_{t}^{\prime}\left[\tilde{\Phi},\tilde{\bar{\phi}}\right]
  =\Gamma_{\Lambda}^{\prime}\left[\Phi,\bar{\phi}\right] \,,
& & \tilde{\Gamma}_{t}\left[\tilde{\Phi},\tilde{\bar{\phi}}\right]=\Gamma_{\Lambda}\left[\Phi,\bar{\phi}\right]\,,
\end{align}
where $t$ is a dimensionless logarithmic scale related to $\Lambda$
via a fixed momentum scale $\mu$ as
\begin{equation}
  \Lambda = \mu\,e^{-t}\,.
\end{equation}

We assume that the cutoff dependence of the cutoff function is given
by
\begin{eqnarray}
\sqrt{R_{\Lambda}}\left[\bar{\phi}\right]\left(x,y\right) & = & \Lambda^{d+1}\sqrt{R}\left[\tilde{\bar{\phi}}\right]\left(\Lambda x,\Lambda y\right)\,,
\end{eqnarray}
and accordingly
\begin{eqnarray}
R_{\Lambda}\left[\bar{\phi}\right]\left(x,y\right) & = &
                                                         \Lambda^{d+2}R\left[\tilde{\bar{\phi}}\right]\left(\Lambda x,\Lambda y\right)\,. \label{eq:RL-Rdimless}
\end{eqnarray}
Both $\sqrt{R}$ and $R$ are dimensionless functions with no explicit
cutoff dependence.  The $\Lambda$-derivative can be rewritten in terms
of the dimensionless cutoff function. For instance, we find
\begin{align}
&\Lambda\partial_{\Lambda}R_{\Lambda}\left[\bar{\phi}\right]\left(x,y\right)
                \nt\\
  & = \Lambda^{d+2}\Biggr[d+2+x\partial_{x}+y\partial_{y}
 -\int_{z}\left(\frac{d-2}{2}+z\partial_{z}\right)\tilde{\bar{\phi}}\left(z\right)\frac{\delta}{\delta\tilde{\bar{\phi}}\left(z\right)}\Biggr]R
    \left[\tilde{\bar{\phi}}\right]\left(\Lambda x,\Lambda y\right)\,.
\end{align}
In the remaining part of this section we work in the dimensionless
framework, and we shall drop the tildes from now on in order not to
burden the notation.

The ERG equations for $S_{t}^{\prime}\left[\sigma,\bar{\phi}\right]$
and $S_{t}\left[\sigma,\bar{\phi}\right]$ are the same and given by
\begin{subequations}
\begin{align}
&\left\{
                \partial_{t}+\int_{x}\left(\frac{d}{2}+x\partial_{x}\right)\sigma\left(x\right)\frac{\delta}{\delta\sigma\left(x\right)}+\int_{x}\left(\frac{d-2}{2}+x\partial_{x}\right)\bar{\phi}\left(x\right)\frac{\delta}{\delta\bar{\phi}\left(x\right)}\right\}
                e^{S_{t}^{\prime}\left[\sigma,\bar{\phi}\right]} \nt\\ 
  & = 
\int_{x,y,z}\Biggr[d+1-\gamma_{t}+x\partial_{x}+z\partial_{z}
    -\int_{w}\left(\frac{d-2}{2}+w\partial_{w}\right)\bar{\phi}\left(w\right)\frac{\delta}{\delta\bar{\phi}\left(w\right)}\Biggr]\sqrt{R}\left(x,z\right)\nt\\
  &\quad \times \sqrt{R}^{-1}\left(z,y\right)
\frac{\delta}{\delta\sigma\left(x\right)}\left\{ \left(\sigma\left(y\right)+\frac{\delta}{\delta\sigma\left(y\right)}\right)e^{S_{t}^{\prime}\left[\sigma,\bar{\phi}\right]}\right\} \,,
\end{align}
and
\begin{align}
&\left\{
                \partial_{t}+\int_{x}\left(\frac{d}{2}+x\partial_{x}\right)\sigma\left(x\right)\frac{\delta}{\delta\sigma\left(x\right)}+\int_{x}\left(\frac{d-2}{2}+x\partial_{x}\right)\bar{\phi}\left(x\right)\frac{\delta}{\delta\bar{\phi}\left(x\right)}\right\} e^{S_{t}\left[\sigma,\bar{\phi}\right]}\nt\\
  & = \int_{x,y,z}\Biggr[d+1-\gamma_{t}+x\partial_{x}+z\partial_{z}
    -\int_{w}\left(\frac{d-2}{2}+w\partial_{w}\right)\bar{\phi}\left(w\right)\frac{\delta}{\delta\bar{\phi}\left(w\right)}\Biggr]\sqrt{R}\left(x,z\right)\nt\\
  &\quad \times \sqrt{R}^{-1}\left(z,y\right)
\frac{\delta}{\delta\sigma\left(x\right)}\left\{ \left(\sigma\left(y\right)+\frac{\delta}{\delta\sigma\left(y\right)}\right)e^{S_{t}\left[\sigma,\bar{\phi}\right]}\right\} \,,
\end{align}
\end{subequations}
where $\gamma_t = \gamma_\Lambda$.  Those for the 1PI actions
$\Gamma_{t}^{\prime}\left[\Phi,\bar{\phi}\right]$ and
$\Gamma_{t}\left[\Phi,\bar{\phi}\right]$ are respectively given by
\begin{subequations}
  \label{eq:dimless-ERG-Gamma}
\begin{align}
&\left\{ \partial_{t}+\int_{x}\left(\frac{d-2}{2}+x\partial_{x}\right)\Phi\left(x\right)\frac{\delta}{\delta\Phi\left(x\right)}+\int_{x}\left(\frac{d-2}{2}+x\partial_{x}\right)\bar{\phi}\left(x\right)\frac{\delta}{\delta\bar{\phi}\left(x\right)}\right\} \Gamma_{t}^{\prime}\left[\Phi,\bar{\phi}\right] \nt\\
  &= - \gamma_t \int_x \Phi (x) \frac{\delta \Gamma'_t [\Phi,
    \bar{\phi}]}{\delta \Phi (x)} +\frac{1}{2}\int_{x,y,z}
    \Biggr[d+2+x\partial_{x}+y\partial_{y}-2\gamma_{t} \nt\\
&\qquad
                                                               -\int_{z}\left(\frac{d-2}{2}+z\partial_{z}\right)\bar{\phi}\left(z\right)\frac{\delta}{\delta\bar{\phi}\left(z\right)}\Biggr]R\left[\bar{\phi}\right]\left(x,y\right)\frac{\delta^{2}W_{t}^{\prime}\left[J,\bar{\phi}\right]}{\delta 
    J\left(x\right)\delta J\left(y\right)}\,,
    \label{eq:dimless-ERGEs-GammaPr}
\end{align}
and
\begin{align}
  &\left\{\partial_{t}+\int_{x}\left(\frac{d-2}{2}+x\partial_{x}\right)\Phi\left(x\right)\frac{\delta}{\delta\Phi\left(x\right)}+\int_{x}\left(\frac{d-2}{2}+x\partial_{x}\right)\bar{\phi}\left(x\right)\frac{\delta}{\delta\bar{\phi}\left(x\right)}\right\} \Gamma_{t}\left[\Phi,\bar{\phi}\right] \nt\\
  & = -\gamma_{t}\int_{x}\Phi\left(x\right)\frac{\delta\Gamma_{t}^{\prime}\left[\Phi,\bar{\phi}\right]}{\delta\Phi\left(x\right)}+\frac{1}{2}\int_{x,y,z}\Biggr[d+2+x\partial_{x}+y\partial_{y}-2\gamma_{t}\nt\\
  &\qquad  -\int_{z}\left(\frac{d-2}{2}+z\partial_{z}\right)\bar{\phi}\left(z\right)\frac{\delta}{\delta\bar{\phi}\left(z\right)}\Biggr]R\left[\bar{\phi}\right]\left(x,y\right)\frac{\delta^{2}W_{t}\left[J,\bar{\phi}\right]}{\delta 
    J\left(x\right)\delta J\left(y\right)}\,,
  \label{eq:dimless-ERGEs-Gamma}
\end{align}
\end{subequations}
where
\begin{subequations}
  \label{eq:dimless-WGamma}
  \begin{align}
    \int_y \frac{\delta^2 W'_t [J, \bar{\phi}]}{\delta J(x) \delta
    J(y)} \left( R [\bar{\phi}] (y, z) - \frac{\delta^2 \Gamma'_t [\Phi,
    \bar{\phi}]}{\delta \Phi (y) \delta \Phi (z)} \right)
    &= \delta (x-z)\,,\\
     \int_y \frac{\delta^2 W_t [J, \bar{\phi}]}{\delta J(x) \delta
    J(y)} \left( R [\bar{\phi}] (y, z) - \frac{\delta^2 \Gamma_t [\Phi,
    \bar{\phi}]}{\delta \Phi (y) \delta \Phi (z)} \right)
    &= \delta (x-z)\,.
  \end{align}
\end{subequations}

Finally, the split Ward identities (\ref{eq:GammaLprime-GammaL-WI})
are transcribed in the dimensionless convention as
\begin{subequations}
  \label{eq:dimless-Gamma-WI}
  \begin{align}
    \frac{\delta}{\delta \bar{\phi} (x)} \Gamma'_t [\Phi, \bar{\phi}]
    &= - \frac{1}{2} \int_{y,z} \frac{\delta R [\bar{\phi}]
      (y,z)}{\delta \bar{\phi} (x)} \frac{\delta^2 W'_t [J,
      \bar{\phi}]}{\delta J(y) \delta J(z)}\,,\label{eq:Gammatprime-WI}\\
    \frac{\delta}{\delta \bar{\phi} (x)} \Gamma_t [\Phi, \bar{\phi}]
    &= z_t \frac{\delta \Gamma_t [\Phi, \bar{\phi}]}{\delta \Phi (x)}
      - \frac{1}{2} \int_{y,z} \frac{\delta R [\bar{\phi}] 
      (y,z)}{\delta \bar{\phi} (x)} \frac{\delta^2 W_t [J,
      \bar{\phi}]}{\delta J(y) \delta J(z)}\,,\label{eq:Gammat-WI}
  \end{align}
\end{subequations}
where $z_t = z_\Lambda$.

\subsection{Fixed points\label{sec:fixed-points}}

The dimensionless framework enables us to discuss how to define fixed
points in the presence of a background field.  For practical
calculations, the 1PI Wilson action $\Gamma$ is the most frequently
used among the three functionals $S, W, \Gamma$ considered in this
work. For this reason, we would like to discuss the ERG equation of
$\Gamma$ and its properties in more detail.

We consider the 1PI Wilson action
$\Gamma_{t}^{\prime}\left[\Phi,\bar{\phi}\right]$ before
$\Gamma_t \left[\Phi,\bar{\phi}\right]$.  
For
$\Gamma'_t \left[\Phi, \bar{\phi}\right]$, the background dependence
comes entirely from the cutoff function $R [\bar{\phi}] (x,y)$.  Since
the cutoff function can depend on virtually an infinite set of
parameters, its dependence on the background is just a particular
dependence.  Whether $\Gamma'_t$ has a fixed point or not is a
physical question, and we expect that the choice of $\bar{\phi}$ is
irrelevant.

To verify this expectation, we need to rewrite the ERG equation
(\ref{eq:dimless-ERGEs-GammaPr}) using the split WI
(\ref{eq:Gammatprime-WI}).  We can obtain
\begin{align}
  &  \lb \partial_t + \int_x \left(\frac{d-2}{2} + \gamma_t + x
    \partial_x \right) \Phi (x) \frac{\delta}{\delta \Phi (x)}\rb
    \Gamma'_t \left[\Phi, \bar{\phi}\right]\nt\\
  &= \frac{1}{2} \int_{x,y,z} \left[
    d+2 + x \partial_x + y \partial_y - 2 \gamma_t \right]
    R[\bar{\phi}] (x,y)
    \frac{\delta^2 W'_t \left[ J, \bar{\phi}\right]}{\delta J(x)
    \delta J(y)}\,.\label{eq:Gammatprime-simpleERG}
\end{align}
This is a differential equation where the background $\bar{\phi} (x)$
is fixed just like a parameter of the cutoff function.  At a fixed
point $\Gamma'_* [\Phi, \bar{\phi}]$, we find
\begin{align}
  &  \int_x \left(\frac{d-2}{2} + \gamma_* + x
    \partial_x \right) \Phi (x) \frac{\delta}{\delta \Phi (x)}
    \Gamma'_* \left[\Phi, \bar{\phi}\right]\nt\\
  &= \frac{1}{2} \int_{x,y,z} \left[
    d+2 + x \partial_x + y \partial_y - 2 \gamma_* \right]
    R[\bar{\phi}] (x,y)
    \frac{\delta^2 W'_* \left[ J, \bar{\phi}\right]}{\delta J(x)
    \delta J(y)}\,.
\end{align}
The background dependence of the fixed point solution is given by the
split WI (\ref{eq:Gammatprime-WI}).

Given a fixed point for $\Gamma'_t [\Phi, \bar{\phi}]$, what about
$\Gamma_t [\Phi, \bar{\phi}]$?  We recall that $\Gamma'_\Lambda$ and
$\Gamma_\Lambda$ are related by (\ref{eq:relation-Gammaprime-Gamma}),
which is transcribed as
\begin{eqnarray}
\Gamma_{t}\left[\Phi,\bar{\phi}\right] & = & \Gamma_{t}^{\prime}\left[\Phi+z_{t}\bar{\phi},\bar{\phi}\right]\,.\label{eq:dimless-rel-Gamma_and_GammaPr}
\end{eqnarray}
This immediately tells us that
$\Gamma_{t}\left[\Phi,\bar{\phi}\right]$ does not display a fixed
point due to the explicit presence of $z_{t}$ on the RHS.  The reason
for this seeming failure of $\Gamma_{t}$ is that, while $\Phi$ is a
renomalized field in the sense that it satisfies a condition such as
$\frac{d}{dp^{2}}\frac{\delta^{2}\Gamma_{t}}{\delta\Phi\delta\Phi}|_{p=0}=1$,
the background field $\bar{\phi}$ is associated with a
non-renormalized field, as suggested by the functional representation
(\ref{eq:functional-repr-W}).

To amend for this disparity between the background and average fields,
let us change the cutoff function.  For clarity we restore the tildes
to denote dimensionless fields.  Instead of (\ref{eq:RL-Rdimless}) and
(\ref{eq:dimless-fields}), we choose
\begin{equation}
  R_\Lambda [\bar{\phi}] (x,y) = \Lambda^{d+2} R \left[
    \tilde{\bar{\phi}}\right] \left( \Lambda x, \Lambda y\right)\,,
\end{equation}
where
\begin{equation}
  \tilde{\bar{\phi}} (x)
  \equiv \Lambda^{-\frac{d-2}{2}} z_\Lambda  \bar{\phi} (x/\Lambda)\,,
\end{equation}
so that
\begin{align}
&  \Lambda \partial_\Lambda R_\Lambda [\bar{\phi}]
                (x,y) \nt\\
  &= \lb d+2 + x \partial_x + y \partial_y
    - \int_z \left( \frac{d-2}{2} + \gamma_\Lambda + z \partial_z \right)
    \tilde{\bar{\phi}} (z) \frac{\delta}{\delta  \tilde{\bar{\phi}}
      (z)} \rb R \left[\tilde{\bar{\phi}}\right] (\Lambda x, \Lambda y)\,.
\end{align}
The dimensional background field $\bar{\phi} (x)$ remains free of an
anomalous dimension, but the dimensionless background field
$\tilde{\bar{\phi}} (x)$ has acquired the anomalous dimension
$\gamma_\Lambda = \gamma_t$.

With the introduction of a new dimensionless background field, let us
revise our previous results.  We redefine
\begin{subequations}
\begin{align}
  \tilde{\Gamma}'_t  [\tilde{\Phi}, \tilde{\bar{\phi}}]
  &=   \Gamma'_\Lambda [\Phi, \bar{\phi}]\,,\\
  \tilde{\Gamma}_t [\tilde{\Phi}, \tilde{\bar{\phi}}]
  &= \Gamma_\Lambda [\Phi, \bar{\phi}]\,.
\end{align}
\end{subequations}
Hence, instead of (\ref{eq:dimless-rel-Gamma_and_GammaPr}), we obtain
\begin{equation}
  \tilde{\Gamma}_t \left[\tilde{\Phi}, \tilde{\bar{\phi}}\right]
  = \Gamma_\Lambda [\Phi, \bar{\phi}] = \Gamma'_\Lambda [\Phi +
  z_\Lambda \bar{\phi}, \bar{\phi}] = \tilde{\Gamma}'_t \left[
    \tilde{\Phi} + \tilde{\bar{\phi}}, \tilde{\bar{\phi}}\right]\,.
\end{equation}
It is already clear that if
$\tilde{\Gamma}'_t [\tilde{\Phi}, \tilde{\bar{\phi}}]$ has a fixed
point for a fixed $\tilde{\bar{\phi}}$, so does
$\tilde{\Gamma}_t [\tilde{\Phi}, \tilde{\bar{\phi}}]$.

Let us revise the ERG equations (\ref{eq:dimless-ERG-Gamma}) and split
WI (\ref{eq:dimless-Gamma-WI}).  With the tildes omitted, the ERG
equations are now given by
\begin{subequations}
  \label{eq:dimless-ERG-Gamma-revised}
\begin{align}
&\left\{ \partial_{t}+\int_{x}\left(\frac{d-2}{2}+\gamma_t
                +x\partial_{x}\right)\Phi\left(x\right)\frac{\delta}{\delta\Phi\left(x\right)}+\int_{x}\left(\frac{d-2}{2}+\gamma_t
                + x\partial_{x}\right)\bar{\phi}\left(x\right)\frac{\delta}{\delta\bar{\phi}\left(x\right)}\right\} \Gamma_{t}^{\prime}\left[\Phi,\bar{\phi}\right] \nt\\
  &= \frac{1}{2}\int_{x,y,z}
    \Biggr[d+2+x\partial_{x}+y\partial_{y}-2\gamma_{t} \nt\\
&\qquad -\int_{z}\left(\frac{d-2}{2}+ \gamma_t
                                                               +z\partial_{z}\right)\bar{\phi}\left(z\right)\frac{\delta}{\delta\bar{\phi}\left(z\right)}\Biggr]R\left[\bar{\phi}\right]\left(x,y\right)\frac{\delta^{2}W_{t}^{\prime}\left[J,\bar{\phi}\right]}{\delta 
    J\left(x\right)\delta J\left(y\right)}\,,
    \label{eq:dimless-ERGEs-GammaPr-revised}
\end{align}
and
\begin{align}
  &\left\{\partial_{t}+\int_{x}\left(\frac{d-2}{2}+\gamma_t+x\partial_{x}\right)\Phi\left(x\right)\frac{\delta}{\delta\Phi\left(x\right)}+\int_{x}\left(\frac{d-2}{2}+\gamma_t+x\partial_{x}\right)\bar{\phi}\left(x\right)\frac{\delta}{\delta\bar{\phi}\left(x\right)}\right\} \Gamma_{t}\left[\Phi,\bar{\phi}\right] \nt\\
  & = \frac{1}{2}\int_{x,y,z}\Biggr[d+2+x\partial_{x}+y\partial_{y}-2\gamma_{t}\nt\\
  &\qquad  -\int_{z}\left(\frac{d-2}{2}+\gamma_t +z\partial_{z}\right)\bar{\phi}\left(z\right)\frac{\delta}{\delta\bar{\phi}\left(z\right)}\Biggr]R\left[\bar{\phi}\right]\left(x,y\right)\frac{\delta^{2}W_{t}\left[J,\bar{\phi}\right]}{\delta 
    J\left(x\right)\delta J\left(y\right)}\,,
  \label{eq:dimless-ERGEs-Gamma-revised}
\end{align}
\end{subequations}
where Eqs.~(\ref{eq:dimless-WGamma}) remain the same.  The split Ward
identities (\ref{eq:dimless-Gamma-WI}) are revised as
\begin{subequations}
  \label{eq:dimless-Gamma-WI-revised}
  \begin{align}
    \frac{\delta}{\delta \bar{\phi} (x)} \Gamma'_t [\Phi, \bar{\phi}]
    &= - \frac{1}{2} \int_{y,z} \frac{\delta R [\bar{\phi}]
      (y,z)}{\delta \bar{\phi} (x)} \frac{\delta^2 W'_t [J,
      \bar{\phi}]}{\delta J(y) \delta J(z)}\,,\label{eq:Gammatprime-WI-revised}\\
    \frac{\delta}{\delta \bar{\phi} (x)} \Gamma_t [\Phi, \bar{\phi}]
    &= \frac{\delta \Gamma_t [\Phi, \bar{\phi}]}{\delta \Phi (x)}
      - \frac{1}{2} \int_{y,z} \frac{\delta R [\bar{\phi}] 
      (y,z)}{\delta \bar{\phi} (x)} \frac{\delta^2 W_t [J,
      \bar{\phi}]}{\delta J(y) \delta J(z)}\,,\label{eq:Gammat-WI-revised}
  \end{align}
\end{subequations}
which are consistent with
\begin{equation}
  \Gamma_t [\Phi, \bar{\phi}] = \Gamma'_t [\Phi + \bar{\phi},
  \bar{\phi}]\,.\label{eq:Gammat-Gammatprime-revised}
\end{equation}

Combining Eqs.~(\ref{eq:dimless-ERGEs-GammaPr-revised}) and
(\ref{eq:Gammatprime-WI-revised}), we reproduce
Eq.~(\ref{eq:Gammatprime-simpleERG}) that we have already discussed.
Now, combining Eqs.~(\ref{eq:dimless-ERGEs-Gamma-revised}) and (\ref{eq:Gammat-WI-revised}), we obtain
\begin{align}
  &\left\{\partial_{t}+\int_{x}\left(\frac{d-2}{2}+\gamma_t+x\partial_{x}\right)
    \left(\Phi\left(x\right) + \bar{\phi} \left(x\right)\right)
    \frac{\delta}{\delta\Phi\left(x\right)} \right\}
     \Gamma_{t}\left[\Phi,\bar{\phi}\right] \nt\\
  & = \frac{1}{2}\int_{x,y,z}\Biggr[d+2+x\partial_{x}+y\partial_{y}-2\gamma_{t}
\Biggr]R\left[\bar{\phi}\right]\left(x,y\right)\frac{\delta^{2}W_{t}\left[J,\bar{\phi}\right]}{\delta 
    J\left(x\right)\delta J\left(y\right)}\,,
    \label{eq:Gammat-simpleERG}
\end{align}
which is consistent with Eq.~(\ref{eq:Gammat-Gammatprime-revised}).
Obviously, if $\Gamma'_t [\Phi, \bar{\phi}]$ has a fixed point for a fixed
$\bar{\phi}$, so does $\Gamma_t [\Phi, \bar{\phi}]$, and vice versa.

We would like to emphasize that we have introduced a single anomalous
dimension, i.e., we did not introduce an anomalous dimension for the
background and another for the average field.  Actually, if one
introduced two independent anomalous dimensions, one could spoil the
actual fixed point.

In the context of the local potential approximation, the role of the
split Ward identity has been discussed in ref.~\cite{Bridle:2013sra},
where a background dependent cutoff was introduced to make it clear
that the realization of the split symmetry Ward identity is absolutely
crucial in order to keep intact the physical content of the system
under study (the Wilson-Fisher fixed point in that case). The present
discussion reiterates this point and makes it very manifest: by means
of the split symmetry Ward identity one can pass from an explicitly
background dependent equation, namely
Eq.~(\ref{eq:dimless-ERGEs-GammaPr}), to
Eq.~(\ref{eq:Gammatprime-simpleERG}), in which the role of the
background is less evident since the latter plays the role of a
parameter of the cutoff.

\section{Explicit checks of the split symmetry \label{sec:checks-of-split-WI}}

In the ERG framework, it should not be taken for granted that any
solution realizes a symmetry: one must check that a solution to the
ERG satisfies the Ward identity associated with the symmetry at hand.
In this section we give two concrete examples and check explicitly
that the Ward identities for background shifts is realized.  The two
examples are the Gaussian fixed point of the real scalar field and the
large $N$ limit of the linear sigma model.
To make the calculations as explicit as possible
and the results easy to read, we limit ourselves to the case of constant background field.

It turns out that in our examples
it is simpler to first calculate $\Gamma^\prime_\Lambda$. 
By using equation (\ref{eq:relation-GammaL-GammaPrimeL}), it is straightforward to obtain $\Gamma_\Lambda$.
Furthermore, we shall take advantage of the simplified form of the ERG
equations discussed in section \ref{sec:fixed-points}, i.e.,
Eq.~(\ref{eq:Gammatprime-simpleERG}), by assuming the validity of the
Ward identity and checking that the latter holds at the very end.

For a constant background field the Ward identity can be written as follows:
\begin{eqnarray}
\partial_{\bar{\phi}}
\Gamma_{\Lambda}^{\prime}\left[\Phi,\bar{\phi}\right]
& = &
-\frac{1}{2}\int_{x,y}
\partial_{\bar{\phi}}R_{\Lambda}\left[\bar{\phi}\right]\left(x,y\right)
\frac{\delta^{2}W_{\Lambda}^{\prime}\left[J,\bar{\phi}\right]}{\delta J\left(x\right)\delta J\left(y\right)}\,,
\label{eq:constant-background-shift-WI-GammaPr}
\end{eqnarray}
where
\begin{equation}
  \int_y \frac{\delta^2 W'_\Lambda [J, \bar{\phi}]}{\delta J(x) \delta
    J(y)} \left( R_\Lambda [\bar{\phi}] (y,z) - \frac{\delta^2
      \Gamma'_\Lambda [\Phi, \bar{\phi}]}{\delta \Phi (y) \delta \Phi
      (z)} \right) = \delta (x-z)\,.
\end{equation}

\subsection{Gaussian fixed point \label{subsec:Gaussian-fixed-point}}

For the Gaussian model, the field dependent part is very simple since
it does not depend on $\Lambda$. However, when working with a background
field, the field independent, but background dependent, part is crucially
important. In the following we focus on calculating such background
dependence and check the realization of the Ward identity associated
with background shifts.

The field independent part of $\Gamma^\prime_\Lambda$ is denoted by by
$U_{t}\left(\bar{\phi}\right)$ and is defined by
$ \int_x U_{t}\left(\bar{\phi}\right) = \Gamma^\prime_\Lambda
\left[0,\bar{\phi}\right]$.
To calculate $U_{t}\left(\bar{\phi}\right)$ we may consider either
Eq.~(\ref{eq:dimless-ERGEs-GammaPr}) or
Eq.~(\ref{eq:Gammatprime-simpleERG}).  It turns out that it is simpler
to work with Eq.(\ref{eq:Gammatprime-simpleERG}).  Let us consider the
Gaussian fixed point, then Eq.~(\ref{eq:Gammatprime-simpleERG}) boils
down to
\begin{eqnarray}
-dU_{*}\left(\bar{\phi}\right) & = & \frac{1}{2}\int_{p}\frac{\left(2-p\partial_{p}\right)R\left[\bar{\phi}\right]\left(p\right)}{p^{2}+R\left[\bar{\phi}\right]\left(p\right)}\,.
\end{eqnarray}
Thus, the fixed point background field potential reads
\begin{eqnarray}
U_{*}\left(\bar{\phi}\right) & = & -\frac{1}{2d}\int_{p}\frac{\left(2-p\partial_{p}\right)R\left[\bar{\phi}\right]\left(p\right)}{p^{2}+R\left[\bar{\phi}\right]\left(p\right)}\,.\label{eq:FP-U_background-1}
\end{eqnarray}

In order to make sure that the background shift Ward identity is satisfied
we must check that
\begin{eqnarray}
\frac{\partial}{\partial\bar{\phi}}U_{*}\left(\bar{\phi}\right) & = & -\frac{1}{2}\int_{p}\frac{\partial R\left[\bar{\phi}\right]\left(p\right)}{\partial\bar{\phi}}\frac{1}{p^{2}+R\left[\bar{\phi}\right]\left(p\right)}\,.\label{eq:background-shift-WI-U_FP-Gaussian}
\end{eqnarray}
To this end let us use the identity
\begin{eqnarray}
-p\partial_{p}\log\frac{p^{2}+R\left(p\right)}{p^{2}} & = & \frac{2R\left(p\right)-p\partial_{p}R\left(p\right)}{p^{2}+R\left(p\right)}\,.
\end{eqnarray}
The above identity allows one to rewrite
\begin{eqnarray}
U_{*}\left(\bar{\phi}\right) & = & -\frac{1}{2}\int_{p}\log\frac{p^{2}+R\left[\bar{\phi}\right]\left(p\right)}{p^{2}}\,.\label{eq:FP-U_background-2}
\end{eqnarray}
Using expression (\ref{eq:FP-U_background-2}) it is straightforward
to check the Ward identity (\ref{eq:constant-background-shift-WI-GammaPr}).

Let us conclude this section by noticing that the solution (\ref{eq:FP-U_background-1})
is a particular solution to Eq.~(\ref{eq:dimless-ERGEs-GammaPr}),
which in this specific case reads as
\begin{eqnarray}
\left(\frac{d-2}{2}\bar{\phi}\partial_{\bar{\phi}}-d\right)U_{*}\left(\bar{\phi}\right) & = & \frac{1}{2}\int_{p}\frac{\left(2-p\partial_{p}-\frac{d-2}{2}\bar{\phi}\partial_{\bar{\phi}}\right)R\left[\bar{\phi}\right]\left(p\right)}{p^{2}+R\left[\bar{\phi}\right]\left(p\right)}\,.
\label{eq:GFP-ERG-background-pot}
\end{eqnarray}
To solve (\ref{eq:GFP-ERG-background-pot}) one could add to the
particular solution (\ref{eq:FP-U_background-1}) a homogeneous
solution: $c\,\bar{\phi}^{\frac{2d}{d-2}}$ with a constant $c$.
However, such a homogeneous solution does not satisfy the Ward identity
(\ref{eq:background-shift-WI-U_FP-Gaussian}).  Hence, one must set
$c=0$.

\subsection{The large $N$ limit}

The ERG equation for the linear sigma model in the large $N$ limit has
been studied extensively
\cite{DAttanasio:1997yph,Morris:1997xj,Blaizot:2005xy,Litim:2018pxe,Sonoda:2023ohb}.
In this section we follow the approach that was developed in
\cite{Morris:1997xj,Sonoda:2023ohb} and references therein and that
was recently used to calculate the vacuum energy density in this limit
\cite{Pagani:2024gdt}.  We work in dimension $d$ with $2<d<4$. Our
main motivation to look into the large $N$ limit is that it offers a
non-perturbative limit of the theory that can be calculated explicitly
and that can be used to check equations in a non-trivial regime. We
are particularly interested in checking the Ward identity.

We consider the Ansatz
\begin{equation}
\Gamma'_{\Lambda}\left[\Phi,\bar{\phi}\right]  =
-\frac{1}{2}\int_{x}\Phi^{i}\left(-\partial^{2}\right)\Phi^{i} + N
\Gamma_{I\Lambda}\left[\frac{\Phi^{i}\Phi^{i}}{2N},\bar{\phi}\right]\,, 
\end{equation}
which is self-consistent to the leading order in large $N$.  We denote
$\varphi\equiv\frac{\Phi^{i}\Phi^{i}}{2N}$, and find it helpful to
introduce the Legendre transform associated with
$\Gamma_{I\Lambda} [\varphi]$:
\begin{equation}
F_{\Lambda}\left[\sigma,\bar{\phi}\right]  =
\Gamma_{I\Lambda}\left[\varphi,\bar{\phi}\right] -
\int_{p}\sigma\left(p\right)\varphi\left(-p\right)\,, 
\end{equation}
where
\begin{equation}
\sigma\left(p\right) \equiv
\frac{\delta\Gamma_{I \Lambda}}{\delta\varphi\left(-p\right)}\left[\varphi,\bar{\phi}\right]\,. 
\end{equation}
It follows that
\begin{eqnarray}
\varphi\left(p\right) & = & -\frac{\delta F_{\Lambda} [\sigma,
                            \bar{\phi}]}{\delta\sigma\left(-p\right)}\,. 
\end{eqnarray}
The ERG equation for $F_{\Lambda}$ is given by
\begin{eqnarray}
-\Lambda\partial_{\Lambda}F_{\Lambda}\left[\sigma,\bar{\phi}\right] &
                                                                      =
  & \frac{1}{2}\int_{p}\Lambda \frac{\partial R_\Lambda [\bar{\phi}]
    (p)}{\partial \Lambda} \cdot \mathcal{G}_{\Lambda;p,-p}\left[\sigma,\bar{\phi}\right]\,, 
\end{eqnarray}
where
\begin{eqnarray}
\int_{q}{\cal G}_{\Lambda;p,-q}\left[\sigma,\bar{\phi}\right]\Biggr\lbrace\left(q^{2}+R_{\Lambda}\left[\bar{\phi}\right]\left(q\right)\right)\delta\left(q-r\right)-\sigma(r-q)\Biggr\rbrace & = & \delta\left(p-r\right)\,.
\end{eqnarray}
We can express $\mathcal{G}_{\Lambda;p,-q}$ as a geometric series in
$\sigma$.  In the large $N$, the background Ward identity is given by
\begin{equation}
  \partial_{\bar{\phi}} F_\Lambda [\sigma, \bar{\phi}]
  = - \frac{1}{2} \int_p \partial_{\bar{\phi}} R_\Lambda [\bar{\phi}]
  (p) \cdot \mathcal{G}_{\Lambda; p, -p} [\sigma, \bar{\phi}]\,.
\end{equation}
This is what we are going to check.

Let us write a generic solution as
\begin{eqnarray}
F_{\Lambda}\left[\sigma,\bar{\phi}\right] & = & F\left[\sigma,\bar{\phi}\right]+I_{\Lambda}\left[\sigma,\bar{\phi}\right]\,,
\end{eqnarray}
where
\begin{eqnarray}
I_{\Lambda}\left[\sigma,\bar{\phi}\right] & \equiv & c_{\Lambda}\left(\bar{\phi}\right)\delta\left(0\right)+c_{1\Lambda}\left(\bar{\phi}\right)\sigma\left(0\right) \nonumber \\
 &  & +\sum_{n=2}^{\infty}\frac{1}{2n}\int_{p_{1}\cdots p_{n}}\sigma\left(p_{1}\right)\cdots\sigma\left(p_{n}\right)\delta\left(\sum_{i=1}^{n}p_{i}\right)I_{n\Lambda}\left(p_{1},\cdots,p_{n};\bar{\phi}\right)
\end{eqnarray}
is a particular solution to the ERG equation, and
$F\left[\sigma,\bar{\phi}\right]$ is an arbitrary functional that is
independent of $\Lambda$ and is determined by imposing a specific
initial condition. We start by discussing the terms of order $n\geq2$
and consider the following ERG equations:
\begin{align}
& -\Lambda\frac{\partial}{\partial\Lambda}I_{n\Lambda}\left(p_{1},\cdots,p_{n};\bar{\phi}\right)  =  \int_{q}f_{\Lambda}\left(q;\bar{\phi}\right)h_{\Lambda}\left(q+p_{1};\bar{\phi}\right)\cdots h_{\Lambda}\left(q+p_{1}\cdots+p_{n-1};\bar{\phi}\right) \nonumber \\
 & \qquad
   +\int_{q}h_{\Lambda}\left(q;\bar{\phi}\right)f_{\Lambda}\left(q+p_{1};\bar{\phi}\right)\cdots
   h_{\Lambda}\left(q+p_{1}\cdots+p_{n-1};\bar{\phi}\right)\nt \\
 &  \qquad
   +\cdots+\int_{q}h_{\Lambda}\left(q;\bar{\phi}\right)h_{\Lambda}\left(q+p_{1};\bar{\phi}\right)\cdots
   f_{\Lambda}\left(q+p_{1}\cdots+p_{n-1};\bar{\phi}\right) \,,
\end{align}
where
\begin{eqnarray}
h_{\Lambda}\left(q;\bar{\phi}\right) & \equiv & \frac{1}{q^{2}+R_{\Lambda}\left[\bar{\phi}\right]\left(q\right)}\,,\\
f_{\Lambda}\left(q;\bar{\phi}\right) & \equiv & \Lambda \frac{\partial
                                                R_{\Lambda}\left[\bar{\phi}\right]\left(q\right)}{\partial
                                                \Lambda} h_{\Lambda}\left(q;\bar{\phi}\right)^{2}\,.
\end{eqnarray}
The ERG for $I_{n\Lambda}$ is solved by
\begin{eqnarray}
I_{n\Lambda}\left(p_{1},\cdots,p_{n};\bar{\phi}\right) & = &
                                                             \int_{q}h_{\Lambda}\left(q;\bar{\phi}\right)h_{\Lambda}\left(q+p_{1};\bar{\phi}\right)\cdots h_{\Lambda}\left(q+p_{1}\cdots+p_{n-1};\bar{\phi}\right)\,. 
\end{eqnarray}
Thus, the terms of order $\sigma^{n}$ in $F_{\Lambda}$ are characterized
by $I_{n\Lambda}$ plus a $\Lambda$-independent term, which we denote
by $F_{n}$. It is straightforward to check that $I_{n\Lambda}$ satisfies
the background shift Ward identity and that for this reason $F_{n}$
must be $\bar{\phi}$-independent. This concludes our discussion for
the terms of order $\sigma^{n}\,(n \ge 2)$ in $F_{\Lambda}$.

To calculate $F_{\Lambda}\left[\sigma=0,\bar{\phi}\right]$ we proceed
via steps similar to those of section \ref{subsec:Gaussian-fixed-point}.
One obtains
\begin{eqnarray}
F_{\Lambda}\left[\sigma=0,\bar{\phi}\right] & = & F\left[0,\bar{\phi}\right]+c_{\Lambda}\left(\bar{\phi}\right)\delta\left(0\right)\,,
\end{eqnarray}
where $\delta (0) = \int_x 1$ is the total space volume, and
\begin{eqnarray}
c_{\Lambda}\left(\bar{\phi}\right) & = & -\frac{1}{2}\int_{q}\log\frac{q^{2}+R_{\Lambda}\left[\bar{\phi}\right]\left(q\right)}{q^{2}}\,.
\end{eqnarray}
Again, the background shift Ward identity requires $F\left[0,\bar{\phi}\right]$
be a constant independent of $\bar{\phi}$.

Finally, let us consider $c_{1\Lambda}\left(\bar{\phi}\right)$, which
satisfies
\begin{eqnarray}
  -\Lambda\partial_{\Lambda}c_{1\Lambda}\left(\bar{\phi}\right) & = &
\frac{1}{2}\int_{q}\Lambda \frac{\partial
R_{\Lambda}\left[\bar{\phi}\right]\left(q\right)}{\partial \Lambda} \,
h_{\Lambda}\left(q,\bar{\phi}\right)^{2}\,.\label{eq:ERG-c1L-largeN}
\end{eqnarray}
A particular solution to (\ref{eq:ERG-c1L-largeN}) is given by
\begin{eqnarray}
c_{1\Lambda}\left(\bar{\phi}\right) & = & 
\frac{1}{2}\int_{q} \left[\frac{1}{q^{2}+R_{\Lambda}\left[\bar{\phi}\right]\left(q\right)}
-\frac{1}{q^2}\right] \,.
\end{eqnarray}
It is straightforward to check that
$c_{1\Lambda}\left(\bar{\phi}\right)$ satisfies the background shift
Ward identity. To obtain the most general term linear in $\sigma$ in
$F_{\Lambda}$ we must add a $\Lambda$-independent term. As has been
shown above, by imposing the Ward identity one deduces that such
$\Lambda$-independent term is also $\bar{\phi}$-independent.

We conclude this section by emphasizing that the large $N$ limit
allowed us to perform a non-trivial check of the background shift Ward
identity.  By imposing this Ward identity one selects a specific
subset of solutions among all the solutions to the ERG equation.

\section{Summary and conclusions} \label{sec:conclusions}

Let us summarize our work,
whose main objective was the study of the background field method in the ERG framework
by focussing on the case of a scalar field theory.

We started out by introducing the Wilson action and the associated modified correlation functions,
which play a key role
in the approach pursued in this work. 
Indeed, assuming that a background field is introduced,
it is by imposing certain desired properties
on the modified correlation functions,
i.e., equation (\ref{eq:imposing-split-no-background-in-cutoffs}),
that one derives the Ward identity that the Wilson
action must satisfy for these properties to be fulfilled.
In dealing with the Wilson action, 
we carefully keep track of field independent, but background dependent, terms.
Perhaps for this reason, the background field method was rarely considered in the Wilson action framework.

Interestingly, one can relate
the Ward identity associated to the Wilson action
to those satisfied by the generating functionals of the connected and the 1PI correlation functions, respectively.
Such Ward identities, 
associated with $W_\Lambda$ and $\Gamma_\Lambda$,
have already been discussed in the literature
\cite{Litim:2002hj,Manrique:2009uh,Manrique:2010mq,Manrique:2010am,Bridle:2013sra,Dietz:2015owa,Labus:2016lkh}.
Our discussion connects these Ward identities to those associated with the Wilson action
showing that they are equivalent.
Moreover, by deriving the Ward identities by imposing suitable properties to the modified correlation function,
one completely circumvents the use of functionals defined by employing a bare action, 
see, e.g., equation (\ref{eq:functional_int_representation_WL}).

The approach presented in this paper
allows one to clarify certain conceptual points.
In particular, in section \ref{sec:anom_dim} we introduced the anomalous dimension
while in section \ref{sec:dimless_ERG} we introduced the dimensionless framework.
Our discussion makes it clear that a single anomalous dimension should be introduced and that it is the
same for both the average field and the background field.
Moreover, in section \ref{sec:fixed-points}, we simplified
the ERG equation by assuming the validity of the Ward identity.
This simplified form makes it explicit that the background is inessential for a scalar theory,
as one expects.

Finally,
in section \ref{sec:checks-of-split-WI},
we considered the Gaussian scalar model fixed point
and the large $N$ limit of the linear sigma model
and verified that split symmetry Ward identity holds.
In our examples, it was simpler to solve first the simplified ERG
equation (\ref{eq:Gammatprime-simpleERG}).
As expected, we saw that a generic solution to the ERG may violate the
Ward identity and a particular solution must be chosen in order for
the Ward identity to be satisfied.

Let us note that the present work lays the ground
for further study regarding the application of the background field method in the ERG framework.
In particular, the case of gauge theories and of
non-linear splits
(e.g., non-linear sigma models)
require an extended discussion.
For instance, in the case of gauge theories we know
that even in the limit $\Lambda\rightarrow 0$
the background does not disappear since the gauge choice depends on the background.

\begin{acknowledgments}
  C.~P.~thanks Kobe University for hospitality where this project was initiated and pursued.
\end{acknowledgments}

\bibliography{paper-background-scalar}

\end{document}